\documentclass[a4paper, journal]{IEEEtran}
\usepackage{color}
\usepackage{amsmath,amsfonts}
\usepackage{algorithm}
\usepackage{algorithmicx}
\usepackage{algpseudocode}
\usepackage{array}
\usepackage{cleveref}
\usepackage[caption=false,font=footnotesize,labelfont=rm,textfont=rm]{subfig}
\usepackage{textcomp}
\usepackage{stfloats}
\usepackage{url}
\usepackage{cite}
\usepackage{verbatim}
\usepackage{graphicx}
\usepackage{caption}
\usepackage{xcolor}
\usepackage{multirow}
\usepackage{multicol}
\usepackage{makecell}
\usepackage{booktabs}
\usepackage{threeparttable}
\usepackage{amsthm}
\theoremstyle{remark}
\newtheorem{remark}{Remark}
\begin{document}

\title{Distribution-Aligned Representation Adaptation for DJSCC over Hybrid Wireless-Wired Networks}

\author{Jiangyuan Guo,~\IEEEmembership{Student Member,~IEEE}, Wei Chen,~\IEEEmembership{Senior Member,~IEEE}, \\ Yuxuan Sun,~\IEEEmembership{Member,~IEEE}, and Bo Ai,~\IEEEmembership{Fellow,~IEEE}
\thanks{Jiangyuan Guo, Wei Chen, Yuxuan Sun and Bo Ai are with the School of Electronic and Information Engineering, Beijing Jiaotong University, Beijing 100044, China. (e-mail: \{jiangyuanguo, weich, yxsun, boai\}@bjtu.edu.cn)}
}

\newcommand{\x} {\boldsymbol{x}}
\newcommand{\y} {\boldsymbol{y}}
\newcommand{\z} {\boldsymbol{z}}
\newcommand{\n} {\boldsymbol{n}}
\newcommand{\s} {\boldsymbol{s}}
\newcommand{\f} {\boldsymbol{f}}
\newcommand{\vect} {\boldsymbol{v}}
\newcommand{\feature} {\boldsymbol{l}}
\newcommand{\bit} {\boldsymbol{b}}
\newcommand{\xhat} {\hat{\boldsymbol{x}}}
\newcommand{\xtilde} {\tilde{\boldsymbol{x}}}
\newcommand{\shat} {\hat{\boldsymbol{s}}}
\newcommand{\yhat} {\hat{\boldsymbol{y}}}
\newcommand{\zhat} {\hat{\boldsymbol{z}}}
\newcommand{\yhatl} {\hat{\boldsymbol{y}}_\lambda}
\newcommand{\yl} {\boldsymbol{y}_\lambda}

\newcommand{\FGM}{\operatorname{FGM}}
\newcommand{\FTM}{\operatorname{FTM}}
\newcommand{\RM}{\operatorname{RM}}
\renewcommand{\algorithmicrequire}{\textbf{Input:}}
\renewcommand{\algorithmicensure}{\textbf{Output:}}
\maketitle

\begin{abstract}
Deep joint source-channel coding (DJSCC) has emerged as a robust alternative to traditional separate coding for communications through wireless channels. Existing DJSCC approaches focus primarily on point-to-point wireless communication scenarios, neglecting end-to-end communication efficiency in hybrid wireless-wired networks such as 5G and 6G communication systems. In such architectures, DJSCC symbols face critical challenges regarding coding efficiency and rate adaptability when forwarded through wired segments. Regarding efficiency, received DJSCC symbols exhibit channel-dependent distributions that differ from the latent feature distributions learned by neural image compression (NIC) models, reducing coding efficiency when these models are directly applied to the received symbols for wired forwarding. In terms of rate adaptability, the system must dynamically regulate its transmission rate to cope with fluctuating network traffic and diverse wired bandwidth constraints. To resolve these challenges, we propose a distribution-aligned representation adaptor (DARA) for transmitting DJSCC symbols over wired backhauls. DARA employs a distribution alignment transform trained through contrastive learning to map noisy received DJSCC symbols into latent features whose distribution is encouraged to align with that of reference features extracted from the original images by a pre-trained NIC model, facilitating efficient entropy coding for wired transmission. Simultaneously, it incorporates a Lagrangian multiplier-based rate control strategy to enable continuous variable-rate coding, closely matching dynamic target rates across hybrid networks. Extensive experiments on diverse datasets demonstrate the superior rate-distortion performance and robustness of DARA compared to existing baselines, validating its potential for wired resource utilization in hybrid transmission scenarios. Specifically, our method achieves peak signal-to-noise ratio (PSNR) gains of up to 0.8dB and 5.1dB compared to neural network-based methods and digital baselines on the CIFAR-10 dataset, respectively. 

\end{abstract}

\begin{IEEEkeywords}
DJSCC, hybrid wireless-wired transmission, distribution mismatch, rate control.
\end{IEEEkeywords}

\section{Introduction}
\IEEEPARstart The rapid evolution of mobile communication technologies from the fifth generation (5G) towards the sixth generation (6G) has significantly transformed how information is generated, transmitted, and consumed. This evolution introduces unprecedented demands on end-to-end (E2E) communication efficiency, particularly across heterogeneous wireless and wired segments. With the explosive growth of intelligent edge devices such as autonomous vehicles \cite{vehicles}, mobile sensors\cite{sensor}, and augmented reality terminals\cite{reality}, the demand for high-throughput, low-latency multimedia communication has become a central concern in modern networks\cite{metaverse}. In particular, the ubiquitous transmission of data over wireless access and the core network has introduced new challenges in communication system design, when harmonizing the varying characteristics of wireless access with the deterministic rate requirements of wired backhauls \cite{Xu2022Attention,10007815}.

To cope with complex wireless conditions, deep joint source-channel coding (DJSCC)\cite{DeepWive, 11219033, digital-sc, 10944429, DVST, 11219364} has emerged as a powerful alternative to traditional separate source and channel coding (SSCC) paradigm. Specifically, DJSCC frameworks exhibit high flexibility, allowing them to be tailored for diverse data modalities (e.g., images \cite{11219033, 11219364}, videos \cite{DeepWive, DVST}) or optimized for specific target tasks, ranging from low-level data reconstruction to high-level semantic analysis, such as image classification \cite{digital-sc} and object detection \cite{10944429}. By replacing this modular architecture with an E2E trainable neural network, DJSCC learns a direct mapping between source signals (e.g., images) and channel symbols, jointly optimizing compression and protection under actual channel conditions \cite{RDJSCC}. A defining merit of DJSCC is its intrinsic adaptability: the encoder can dynamically shape the latent representations to match the instantaneous wireless channel state, thereby avoiding the \emph{cliff effect} and achieving graceful performance degradation even under severe noise \cite{D2JSCC,10786312}.

Despite these advantages, most existing DJSCC approaches assume fully wireless point-to-point communication scenarios\cite{VideoQA-SC,DeepWive,NTSCC}. In practice, however, E2E communication often spans \emph{hybrid network segments}. Source data is typically captured by edge devices and transmitted via wireless access links to a gateway, followed by wired backhaul for long-distance delivery. This architecture is prevalent in scenarios such as multi-hop relay transmission\cite{Bian_Hybird} and mobile cloud offloading\cite{10945654} where a mobile device offloads a heavy computational task to a distant cloud server. In these hybrid settings, the core challenge shifts from solely combating wireless noise to managing the efficient transition of DJSCC symbols across diverse link protocols\cite{Bian_JSAC}.

\begin{figure}[!t]
\centering
\includegraphics[width=0.35\textwidth]{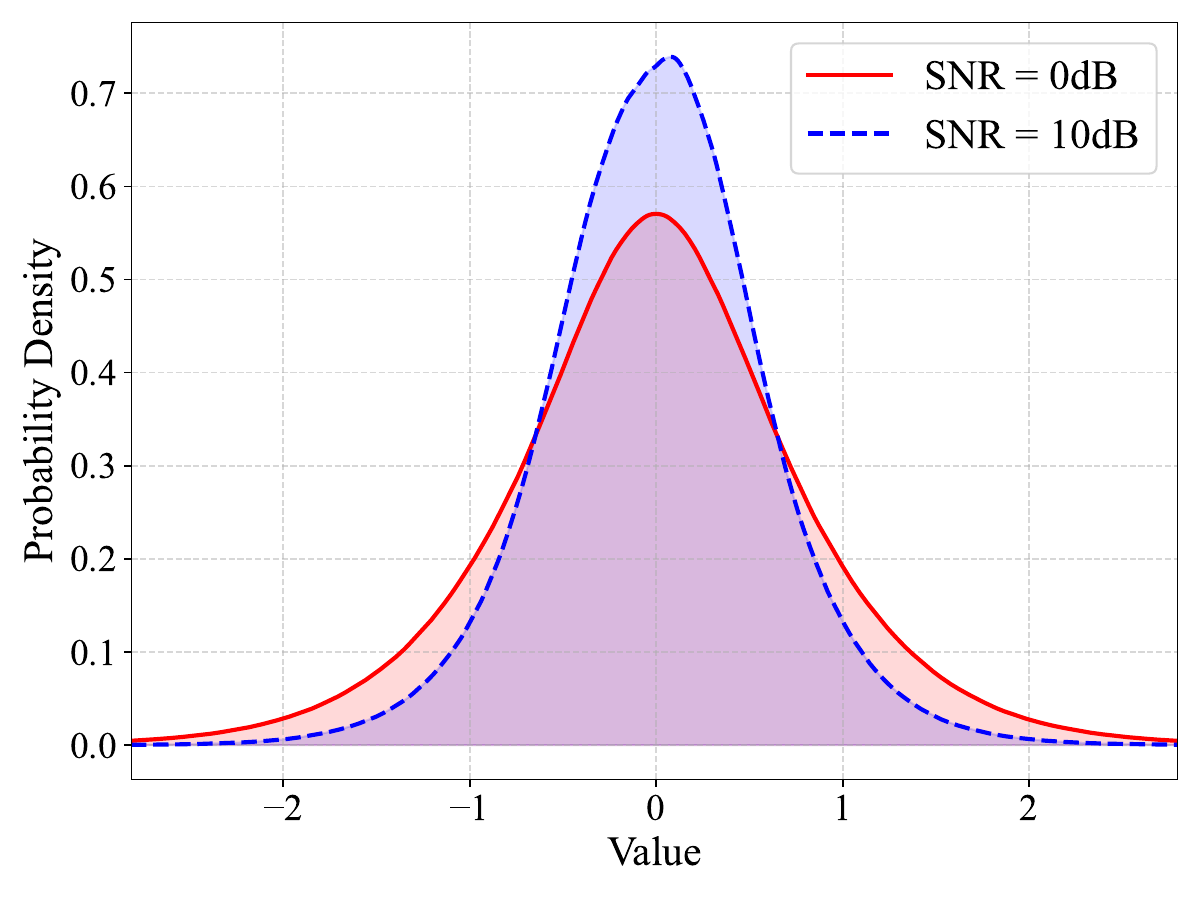}
\caption{Visualization of distribution of DJSCC symbols under different channel SNRs on the CIFAR-10 dataset.}
\label{Fig:symbol_pdf}
\end{figure}

First, a fundamental challenge regarding \emph{coding efficiency} arises from the distribution mismatch between received DJSCC symbols and the latent features learned by neural image compression (NIC) models. DJSCC representations are optimized for wireless robustness, and their distributions vary with channel conditions. As illustrated in Fig. \ref{Fig:symbol_pdf}, Gaussian kernel
density estimation (KDE) on the CIFAR-10 dataset shows that DJSCC symbol distributions vary with wireless signal-to-noise ratio (SNR), exhibiting broader distributions at low SNRs and more concentrated distributions at high SNRs. NIC entropy models, however, are learned from image-derived latent features, which remain unchanged for a given image. Thus, even for the same source image, the distribution of received DJSCC symbols changes with wireless SNR, whereas the latent features extracted by a fixed NIC encoder remain unchanged, which results in a distribution mismatch and reduces coding efficiency when these models are directly applied to wired forwarding. This motivates a \emph{distribution alignment} mechanism that transforms received DJSCC symbols into features whose distribution
aligns with the latent feature distribution of a pre-trained NIC encoder, facilitating efficient wired transmission.

Second, the dynamic nature of wired backhauls imposes varying transmission rate constraints across different network states, demanding high \emph{rate adaptability}. This necessitates \emph{a controllable variable-rate coding scheme} that can adaptively adjust the transmission rate of DJSCC symbols to \emph{align with the current bandwidth budget}, thereby preventing network congestion while guaranteeing reconstruction quality. Recent advances in neural compression have explored variable-rate coding through gain-based modules \cite{Li_2023_CVPR,Li_2024_CVPR} or auxiliary condition inputs, such as quantization steps \cite{10222717} and rate-distortion (R-D) hyperparameters \cite{9998500}. However, these methods are predominantly designed for the pixel domain, where inputs follow relatively stable natural image or video distributions. In contrast, DJSCC symbols possess a unique structure governed by fluctuating wireless channel conditions, as demonstrated in our KDE analysis. Such distinct data manifolds and volatile distributions render the direct application of existing strategies suboptimal, requiring an approach to achieve joint distribution normalization and rate adaptation.

Recent works have explored DJSCC design within hybrid multi-hop scenarios to bridge the gap between diverse network segments. For instance, \cite{Bian_JSAC} introduced a gain-based approach to adjust coding behavior. However, its reliance on a predefined, discrete set of gain vectors limits the system to a few fixed rate options and struggles to provide fine-grained alignment with varying transmission rate constraints. Alternatively, a learnable quantization method was proposed in \cite{digital-sc} to adaptively determine quantization intervals for effectively discretizing DJSCC symbols. Nevertheless, this approach fails to utilize entropy coding, which inevitably leads to severe coding inefficiency.

In this work, we focus on efficient image transmission in E2E DJSCC systems operating over \emph{hybrid wireless-wired} networks. To address the critical challenges of transmission efficiency and rate adaptability, we propose \emph{distribution-aligned representation adaptor} (DARA) to seamlessly forward wireless-optimized DJSCC symbols through wired segments. To address the distribution mismatch, we introduce a \emph{distribution alignment transform} (DAT) that transforms noisy received DJSCC symbols into latent features for wired coding. Through contrastive supervision, DAT aligns the distribution of its output features with that of reference features extracted from the original images by a pre-trained NIC encoder. These channel-independent references guide DAT to reduce channel-induced variations while preserving source information, improving entropy coding efficiency for wired forwarding. Furthermore, to enhance rate adaptability, DARA incorporates a \emph{Lagrangian multiplier predictor} (LMP) to predict an appropriate Lagrangian multiplier for each target rate, thereby achieving continuous variable-rate coding that closely aligns with the dynamic transmission rate constraints of wired backhauls. Importantly, our adaptor is compatible with various existing DJSCC architectures without requiring modifications to the base encoder-decoder pipeline, making it a practical solution for DJSCC-based hybrid transmission systems.

The main contributions of this paper are:
\begin{enumerate}
    \item We develop DARA to adapt DJSCC symbols for transmission over wired backhauls with varying rate constraints. Furthermore, this adaptation is achieved without requiring structural modifications to the base DJSCC architecture, enhancing the flexibility and E2E performance of DJSCC in hybrid wireless-wired scenarios.
    \item We introduce a DAT to align the distribution of features derived from channel-dependent received DJSCC symbols with the latent feature distribution of a pre-trained NIC encoder. Specifically, we train DAT through contrastive supervision using reference features extracted from the original images by the NIC encoder. These channel-independent references guide DAT to produce coding-efficient representations under varying wireless conditions, improving entropy coding efficiency for wired forwarding.
    \item
    We develop a $\lambda$-conditioned feature coding network
    that supports continuous rate adaptation within a single model.
    To meet varying wired rate constraints, we further introduce
    a Lagrangian multiplier predictor that selects an appropriate
    $\lambda$ based on the aligned features and target rate.
    Together, these modules enable flexible adjustment of the
    rate-distortion (R-D) trade-off and accurate rate control without repeated
    trial encoding.
    \item Experiments on datasets with different image resolutions demonstrate the effectiveness of DARA and highlight its superiority over existing baselines in hybrid transmission settings. Specifically, results on CIFAR-10 dataset show peak signal-to-noise ratio (PSNR) gains of up to 0.8dB and 5.1dB compared to existing neural network-based methods \cite{digital-sc, Bian_JSAC} and digital baselines, respectively.
    
\end{enumerate}

The rest of this paper is organized as follows. The system model and preliminaries are introduced in \Cref{Section:system model} and \Cref{Section:preliminary}, respectively. We elaborate our proposed methods and the detailed network architectures in \Cref{Section:proposed_methods}. Furthermore, \Cref{Section:numerical_results} provides the quantified numerical results and the corresponding analysis. Finally, \Cref{Section:conclusion} summarizes this paper and gives conclusions.

\textit{Notations}: In this paper, lowercase letters, e.g., ${x}$, denote scalars. Bold lowercase letters, e.g., $\x$, denote vectors or tensors. $\boldsymbol{I}$ denotes the identity matrix. $\mathcal{CN}(\mu,\sigma^{2})$ and $\mathcal{N}(\mu,\sigma^{2})$ denote the complex Gaussian distribution and the real Gaussian distribution with mean $\mu$ and variance $\sigma^{2}$, respectively. $\mathcal{U}(a,b)$ denotes the uniform distribution with the start $a$ and end $b$. $\mathbb{R}$ and $\mathbb{C}$ denote the real set and the complex set, respectively. $\mathbb{E}[\cdot]$ denotes the statistical expectation operation. Given $\x$, $p(\x)$ represents its probability distribution and $p(\x|\y)$ represents its conditional probability distribution conditioned on $\y$.

\begin{figure*}[!t]
\centering
\includegraphics[width=1\textwidth]{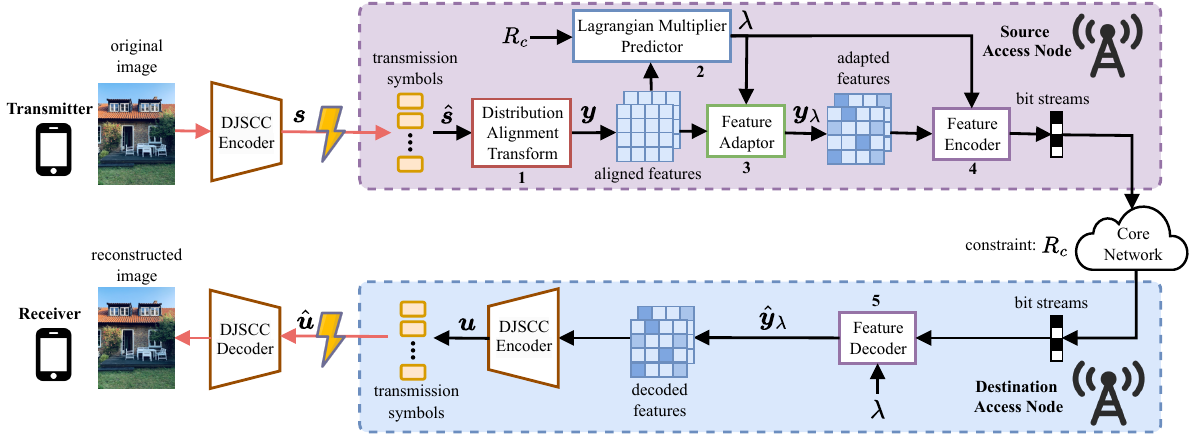}
\caption{DARA-DJSCC for E2E transmission over hybrid
wireless-wired networks. DARA integrates distribution
alignment, Lagrangian multiplier prediction, feature
adaptation, and feature encoding and decoding for
efficient wired forwarding.}
\label{Fig:pipeline}
\end{figure*}

\section{System Model} \label{Section:system model}

In this section, we introduce the DARA-DJSCC transmission
scheme for hybrid wireless-wired networks, describe its
processing pipeline, and formulate the optimization objective.

As illustrated in Fig.~\ref{Fig:pipeline}, we consider E2E
image transmission between two edge devices connected through
independent wireless access links and a reliable wired core
network with a transmission rate constraint $R_c$.
To support efficient wired forwarding of received DJSCC
symbols, DARA first aligns their distributions and then
uses a feature encoder-decoder pair to convert the aligned
features into bitstreams and recover them at the destination
access node. A feature adaptor conditioned on the Lagrangian
multiplier adjusts the coding rate within a single model,
while a Lagrangian multiplier predictor selects the multiplier
according to $R_c$ and the aligned features.
All learnable modules are implemented by DNNs.

Given an original image $\x\in\mathbb{R}^{3\times{H}\times{W}}$ with height $H$ and width $W$, the transmitter encodes $\x$ into transmission symbols $\s\in\mathbb{C}^{n_1}$ by a DJSCC encoder parameterized by $\boldsymbol{\theta}$, and then transmits them to a neighboring access node (source access node) via a wireless access channel.

We consider AWGN and Rayleigh block fading channels,
with received symbols $\shat\in\mathbb{C}^{n_1}$ given by
\begin{equation}
    \shat=h\s+\n,
\end{equation}
where $\n\sim\mathcal{CN}(0,\sigma_n^2\boldsymbol{I})$
and $\sigma_n^2$ denotes the average noise power.
For Rayleigh block fading, $h\sim\mathcal{CN}(0,1)$
remains constant within each transmitted symbol block.
The AWGN channel is obtained as the special case $h=1$. Note that for Rayleigh block fading channels, minimum mean squared error (MMSE) equalization is applied to $\shat$ to mitigate fading. For the considered AWGN and Rayleigh block fading channels, with unit average transmit symbol power, the average
wireless SNR is given by $\eta=1/\sigma_n^2$.

At the source access node, DAT transforms the received symbols
$\shat$ into aligned features
$\y\in\mathbb{R}^{c\times(H/d)\times(W/d)}$:
\begin{equation}
    \y=\operatorname{DAT}(\shat,\eta_1;\boldsymbol{\tau}),
\end{equation}
where $d$ is the downsampling factor and $\eta_1$ denotes SNR of the source access channel. These features retain
information for image reconstruction while aligning with
compression-based distribution priors. Given $\y$ and the
wired rate constraint $R_c$, LMP predicts the Lagrangian
multiplier $\lambda$:
\begin{equation}
    \lambda=\operatorname{LMP}(\y,R_c;\boldsymbol{\xi}).
    \label{Eq:prediction}
\end{equation}
The predicted $\lambda$ guides feature adaptation and encoding:
\begin{equation}
    \boldsymbol{y}_{\lambda}
    =\operatorname{FA}(\y,\lambda;\boldsymbol{\omega}),
    \label{Eq:transformation}
\end{equation}
\begin{equation}
    \boldsymbol{b}
    =\operatorname{FE}(\boldsymbol{y}_{\lambda},
    \lambda;\boldsymbol{\omega}),
\end{equation}
where $\boldsymbol{y}_{\lambda}$ has the same dimensions as
$\y$, and $\boldsymbol{b}$ is the resulting bitstream.
The source access node forwards $\boldsymbol{b}$ together
with $\lambda$, represented as a 16-bit half-precision
floating-point number, through the wired core network.

At the destination access node, the feature decoder recovers
the adapted features using the received bitstream and $\lambda$:
\begin{equation}
    \hat{\boldsymbol{y}}_{\lambda}
    =\operatorname{FD}(\boldsymbol{b},
    \lambda;\boldsymbol{\omega}).
\end{equation}
The recovered features are re-encoded into DJSCC symbols
$\boldsymbol{u}\in\mathbb{C}^{n_2}$ and transmitted over
the destination wireless access link. The receiver then
reconstructs $\xhat$ from the noisy symbols
$\hat{\boldsymbol{u}}$ using the DJSCC decoder.
If the receiver resides in the core network, a reconstruction
network directly recovers $\xhat$ from
$\hat{\boldsymbol{y}}_{\lambda}$.
Here, $\boldsymbol{\tau}$, $\boldsymbol{\xi}$,
$\boldsymbol{\omega}$, and $\boldsymbol{\theta}$ denote
the learnable parameters of DAT, LMP, the feature adaptor
and encoder-decoder, and the DJSCC modules, respectively.

We define the source and destination wireless channel bandwidth
ratios (CBRs) as $\rho_s=n_1/(3HW)$ and $\rho_d=n_2/(3HW)$,
respectively, and the wired transmission rate as
$R=\operatorname{len}(\bit)/(HW)$ in bits per pixel (BPP).
The learnable parameters $\boldsymbol{\theta}$,
$\boldsymbol{\tau}$, $\boldsymbol{\omega}$, and
$\boldsymbol{\xi}$ correspond to the DJSCC modules, DAT,
feature adaptor and encoder-decoder, and LMP, respectively.
The system minimizes reconstruction distortion subject to
wireless bandwidth and wired rate constraints:
\begin{subequations}
\begin{align}
\min_{\boldsymbol{\theta},\boldsymbol{\tau},
     \boldsymbol{\omega},\boldsymbol{\xi}}
&\quad d(\x,\xhat)\\
\mathrm{s.t.}\ &\quad
\rho_s\le\rho_1,\quad \rho_d\le\rho_2,\quad R\le R_c,
\end{align}
\label{Eq:overall}
\end{subequations}
where $\rho_1$ and $\rho_2$ are the wireless CBR limits,
and $R_c$ is the wired rate limit. The distortion measure
$d(\cdot,\cdot)$ can be MSE or a perceptual metric such as
LPIPS~\cite{Zhang_2018_CVPR}; we use MSE in the main experiments.
Algorithm~\ref{alg:forward_pipeline} summarizes the forward
processing pipeline.

\begin{algorithm}[!t]
\caption{Forward processing pipeline of the proposed DARA-DJSCC framework.}
\label{alg:forward_pipeline}
\begin{algorithmic}[1]
\Require Original image $\x$; wireless channel gains $h_1, h_2$; SNR of source and destination access channels $\eta_1, \eta_2$; target wired transmission rate constraint $R_c$.
\Ensure Reconstructed image $\xhat$.
\Statex \textbf{// Stage 1: Transmitter Encoding}
\State $\s = \operatorname{DJSCC-Encoder}(\x, \eta_1; \boldsymbol{\theta})$ \Comment{Initial DJSCC encoding}
\State $\shat = W_1(\s, h_1, \eta_1)$ \Comment{Transmission through source access channel}
\Statex \textbf{// Stage 2: Source Access Node Processing}
\State $\y = \operatorname{DAT}(\shat, \eta_1; \boldsymbol{\tau})$ \Comment{Distribution alignment}
\State $\lambda = \operatorname{LMP}(\y, R_c; \boldsymbol{\xi})$ \Comment{$\lambda$ prediction for target-rate control}
\State $\y_\lambda = \operatorname{FA}(\y, \lambda; \boldsymbol{\omega})$ \Comment{Feature adaptation based on $\lambda$ }
\State $\boldsymbol{b} = \operatorname{FE}(\y_\lambda, \lambda; \boldsymbol{\omega})$ \Comment{Feature encoding into bitstream }
\Statex \textbf{// Stage 3: Wired Backhaul Forwarding}
\State Forward $\boldsymbol{b}$ and $\lambda$ through the core network.
\Statex \textbf{// Stage 4: Destination Access Node Processing}
\State $\hat{\y}_\lambda = \operatorname{FD}(\boldsymbol{b}, \lambda; \boldsymbol{\omega})$ \Comment{Feature decoding at destination node }
\State $\boldsymbol{u} = \operatorname{DJSCC-Encoder}(\hat{\y}_\lambda, \eta_2; \boldsymbol{\theta})$ \Comment{DJSCC encoding for destination access channel }
\State $\hat{\boldsymbol{u}} = W_2(\boldsymbol{u}, h_2, \eta_2)$ \Comment{Transmission through destination access channel }
\Statex \textbf{// Stage 5: Receiver Decoding}
\State $\xhat = \operatorname{DJSCC-Decoder}(\hat{\boldsymbol{u}}, \eta_2; \boldsymbol{\theta})$ \Comment{DJSCC decoding \& Final image reconstruction }\State \Return $\xhat$.\end{algorithmic}
\end{algorithm}

To handle the wired transmission rate constraint $R \le R_c$, we introduce a Lagrangian multiplier and approximately convert the constrained optimization into the following standard rate-distortion (R-D) objective:
\begin{equation}
\label{Eq:lagrangian}
\min_{\boldsymbol{\theta},\boldsymbol{\tau},\boldsymbol{\omega},\boldsymbol{\xi}}
\lambda \cdot d(\x,\xhat) + R,
\end{equation}
where $\lambda$ controls the trade-off between reconstruction distortion and transmission rate. By adjusting $\lambda$, the model can operate at different points on the R-D curve and adapt its coding rate to different wired transmission constraints.

\section{Preliminaries} \label{Section:preliminary}
Recently, neural image compression (NIC) has achieved coding performance comparable to or exceeding that of advanced conventional frameworks like joint photographic experts group (JPEG) or better portable graphics (BPG).

\begin{figure}[!t]
\centering
\includegraphics[width=0.5\textwidth]{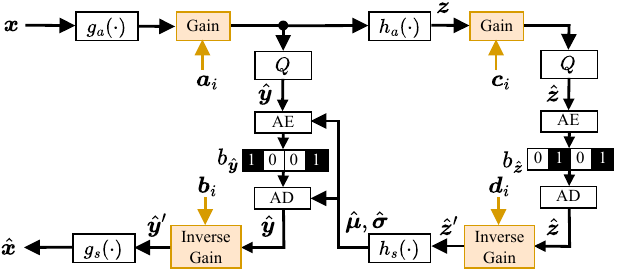}
\caption{The variable-rate neural image compression framework. The colored blocks indicate the modules introduced to support different rates.}
\label{Fig:NIC}
\end{figure}

Classical NIC frameworks are illustrated in Fig. \ref{Fig:NIC} without colored blocks, where the encoding and decoding of an image $\x$ are formulated as:
\begin{align}
\begin{aligned}
    &\operatorname{Encoding}: \y = g_a(\x),\yhat=Q(\y),\bit_{\yhat}=\operatorname{AE}(\yhat|p({\yhat})), \\
    &\operatorname{Decoding}:\yhat=\operatorname{AD}(\bit_{\yhat}|p({\yhat})),\xhat=g_s(\yhat), \\
\end{aligned}
\end{align}
where $g_a(\cdot)$ and $g_s(\cdot)$ are the encoder and decoder parameterized by neural networks, respectively. $\y$, $\yhat$, $\bit_{\yhat}$ and $\xhat$ denote the feature, quantized feature, bit stream and reconstructed image, respectively. $Q(\cdot)$, $\operatorname{AE}(\cdot)$, and $\operatorname{AD}(\cdot)$ are quantization, arithmetic encoding and decoding functions, respectively. $p({\yhat})$ is estimated probability distribution of $\yhat$.

A key component of NIC is accurate estimation of the unknown true probability distribution of $\y$, since the coding rate is theoretically equal to the cross-entropy between estimated distribution $p(\yhat)$ and the actual distribution. To this end, another latent vector $\z$ is extracted from $\y$ and serves as the hyperprior for entropy estimation. With the help of $\z$, dependencies in $\yhat$ can be effectively removed and each element $\hat{y}_i$ is modeled by a Gaussian distribution
$\mathcal{N}(\hat{\mu}_i,\hat{\sigma}_i^2)$, where
$\hat{\mu}_i$ and $\hat{\sigma}_i$ denote its mean and
standard deviation, respectively. This process can be formulated as follows:
\begin{align}
\begin{aligned}
&\z = h_a(\y), \ \zhat=Q(\z),\ \hat{\boldsymbol{\mu}},\hat{\boldsymbol{\sigma}}=h_s(\zhat), \\
&\bit_{\yhat}=\operatorname{AE}(\yhat \mid \hat{\boldsymbol{\mu}},  \hat{\boldsymbol{\sigma}}), \ \yhat=\operatorname{AD}(\bit_{\yhat} \mid \hat{\boldsymbol{\mu}}, \hat{\boldsymbol{\sigma}}), \\
\end{aligned}
\end{align}
where $h_a(\cdot)$, $h_s(\cdot)$ are the hyperencoder and hyperdecoder parameterized by neural networks. $\hat{\boldsymbol{\mu}}$ and $\hat{\boldsymbol{\sigma}}$ represent the vectors composed of all $\hat{\mu}_i$ and $\hat{\sigma}_i$, respectively.

Since the quantization operation $Q(\cdot)$ rounds each element $y_i$ in $\y$ to the nearest integer, its non-differentiability prevents direct gradient-based E2E training. Therefore, a ``quantization noise'' $\n^\prime\sim \mathcal{U}(-\frac{1}{2},\frac{1}{2})$ is added to $\y$ and $\z$, forming proxies $\tilde{\y}$ and $\tilde{\z}$ to simulate $\yhat$ and $\zhat$ during the training for E2E optimization. Finally, the conditional distribution $p({\yhat\mid\zhat})$ is relaxed to:
\begin{align}
\begin{aligned}
    p(\tilde{\y} \mid \tilde{\boldsymbol{z}})=&\prod_i\left(\mathcal{N}\left(\tilde{\mu}_i, \tilde{\sigma}_i^2\right) * \mathcal{U}\left(-\frac{1}{2}, \frac{1}{2}\right)\right)\left(\tilde{y}_i\right) \\
\end{aligned}
\label{Eq:py}
\end{align}
where $[\tilde{\boldsymbol{\mu}},\tilde{\boldsymbol{\sigma}}]=h_s(\tilde{\boldsymbol z})$ collects the estimated means and standard deviations. As for $p(\zhat)$, it is modeled by a non-parametric, fully factorized density model with relaxation:
\begin{equation}
p_{\tilde{\boldsymbol{z}}|\boldsymbol{\psi}}
(\tilde{\boldsymbol{z}}|\boldsymbol{\psi})
=
\prod_i
\left[
p_{z_i|\psi_i}(\cdot|\psi_i)
*
\mathcal{U}\left(-\frac12,\frac12\right)
\right](\tilde{z}_i).
\end{equation}
where $\boldsymbol{\psi}$ collects the parameters of the factorized density model.

The optimization objective of NIC is R-D performance, i.e., the trade-off between rate cost and reconstruction quality:
\begin{equation}
    \mathcal{L}=\lambda d(\x,\xhat)+R(\bit_{\yhat},\bit_{\zhat}),
\label{Eq:RD}
\end{equation}
where $\lambda$ controls the trade-off between the two terms. $d(\cdot,\cdot)$ denotes distortion measurements, and $R(\cdot)$ denotes rates of corresponding bit streams, which can be approximately expressed as the entropy of $\hat{\y}$ and $\hat{\z}$:
\begin{equation}
    R(\bit_{\yhat},\bit_{\zhat})=\frac{1}{HW}\mathbb{E}(-\log_2{p(\yhat\mid\zhat)}-\log_2{p(\zhat)}).
\end{equation}
Again, $\tilde{\y}$ and $\tilde{\z}$ are used to replace $\yhat$ and $\zhat$ to measure rate costs during training, respectively. The coding rate is measured in bits per pixel (BPP).

\begin{remark}\emph{(Differences between Coding of DJSCC Symbols and Images):}
Although Eq. \eqref{Eq:overall} is similar to the objective of NIC, directly applying NIC to DJSCC symbols is ineffective. Unlike natural image features with relatively stable distributions, DJSCC symbols vary with wireless channel conditions, as shown in Fig. \ref{Fig:symbol_pdf}. Moreover, DJSCC symbols jointly encode source information and channel-adaptive characteristics, resulting in a latent manifold different from that of conventional image features. Therefore, directly entropy-coding these symbols with standard NIC models leads to distribution mismatch and degraded coding efficiency.
\label{remark:mismatch}
\end{remark}

\begin{figure}[!t]
\centering
\includegraphics[width=0.48\textwidth]{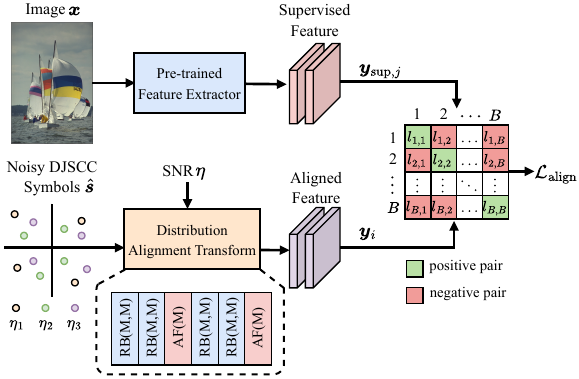}
\caption{Supervised contrastive learning for distribution
alignment. RB denotes a residual block with $3\times3$ convolutions
and GELU activation, and AF denotes the attention block
in \cite{Xu2022Attention}.
The channel dimension is $M=256$.}
\label{Fig:Contrastive_Learning}
\end{figure}

\section{Proposed Methods} \label{Section:proposed_methods}
In this section, we first introduce distribution alignment
for DJSCC symbols under varying wireless channel conditions,
followed by continuous variable-rate coding of the aligned
features. We then present Lagrangian multiplier-based rate
control, extend the framework to high-resolution images,
and describe the training procedure.

\subsection{Distribution Alignment via Supervised Contrastive Learning}
In this section, we introduce a DAT trained through supervised contrastive learning to align the distribution of features derived from received DJSCC symbols with that of reference features extracted by a pre-trained NIC encoder, enabling efficient wired transmission.

As discussed in \Cref{remark:mismatch}, DJSCC symbols jointly
represent source information and channel-adaptive characteristics,
resulting in distributions that differ from those of conventional
NIC features. Directly applying NIC to these symbols can
therefore lead to distribution mismatch and reduced coding
efficiency. To mitigate this mismatch, we use latent features
from a pre-trained NIC model as reference representations
to guide the training of DAT.

Specifically, we adopt the feature extractor of the pre-trained
NIC model in \cite{balle2018variational} as a teacher network,
whose parameters remain fixed during DAT training.
For each image $\x$, the teacher produces a reference feature
$\y_{\text{sup}}$, while DAT maps the corresponding noisy
DJSCC symbols $\shat$ to a feature $\y$.
These two representations share the same source content
but are obtained through different paths: the teacher
processes the original image directly, whereas DAT operates
on symbols affected by wireless transmission.
The teacher feature thus serves as a channel-independent
reference, guiding DAT to preserve source information
while adapting its output to the NIC latent space.

As illustrated in Fig.~\ref{Fig:Contrastive_Learning},
the DAT feature and teacher feature derived from the same
image form a positive pair, whereas features derived from
different images form negative pairs. Treating the feature
representations as vectors, the similarity between the
$i$-th DAT feature $\y_i$ and the $j$-th teacher feature
$\y_{\text{sup},j}$ is defined as
\begin{equation}
    \ell_{i,j}
    =\frac{\y_i^\top\y_{\text{sup},j}}
    {\|\y_i\|_2\|\y_{\text{sup},j}\|_2
    \cdot\tau_{\text{temp}}},
\end{equation}
where $\|\cdot\|_2$ denotes the $\ell_2$ norm and
$\tau_{\text{temp}}$ is a temperature parameter controlling
the sharpness of the similarity-based matching probabilities.

Based on these similarities, we employ a symmetric
cross-entropy loss to establish bidirectional correspondence
between the DAT and teacher representations. For a mini-batch
of $B$ samples, the alignment loss is
\begin{equation}
\label{eq:align_loss}
\mathcal{L}_{\text{align}}
=\frac{1}{2B}\sum_{i=1}^{B}
\left(
-\log\frac{e^{\ell_{i,i}}}
{\sum_{j=1}^{B}e^{\ell_{i,j}}}
-\log\frac{e^{\ell_{i,i}}}
{\sum_{j=1}^{B}e^{\ell_{j,i}}}
\right).
\end{equation}
Here, $\ell_{i,i}$ denotes the similarity of a positive pair,
while $\ell_{i,j}$ and $\ell_{j,i}$ for $i\neq j$ correspond
to negative pairs. The first term increases the similarity of each DAT feature
to its paired teacher feature relative to unpaired teacher
features in the mini-batch. The second term applies the same
objective in the reverse direction, jointly encouraging
bidirectional alignment between the two feature spaces.

Unlike point-wise feature alignment, which penalizes
element-wise differences, this contrastive objective
supervises the relative similarities between representations.
It encourages DAT features to match their source-specific
teacher references while remaining distinguishable from
representations of other images. This allows DAT to learn
content-aware correspondence with the NIC latent space
without requiring exact element-wise agreement with the
teacher features.

The resulting supervision provides a consistent reference
for adapting DJSCC symbols across wireless channel conditions.
By encouraging representations of the same source to remain
close to their channel-independent teacher features, DAT
can reduce channel-induced variations while preserving
information needed for reconstruction. Although the contrastive
loss does not explicitly match probability distributions,
it guides the learned representations toward a latent space
designed for compression, helping mitigate distribution
mismatch and improve entropy coding efficiency for wired
forwarding.

\begin{figure}[t]
\centering
\subfloat[Coding block (CB).]{\includegraphics[width=0.34\linewidth]{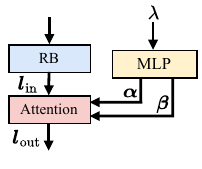}%
\label{Fig:FA_FE_a}}
\subfloat[Feature adaptor.]{\includegraphics[width=0.3\linewidth]{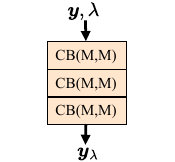}%
\label{Fig:FA_FE_b}}
\subfloat[Feature encoder.]{\includegraphics[width=0.35\linewidth]{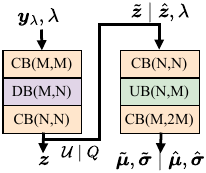}%
\label{Fig:FA_FE_c}}
\caption{Attention-based variable-rate feature coding. CB, DB, and UB denote coding, downsampling, and upsampling blocks, respectively; channel dimensions are given in parentheses
($M=256$, $N=192$).}
\label{Fig:FA_FE}
\end{figure}

\subsection{Continuous Variable-Rate Feature Coding} \label{Section:variable-rate}
In this section, we introduce a $\lambda$-conditioned feature
coding method that enables continuous rate adaptation within
a single model.

After distribution alignment, the features $\y$ are encoded
into a bitstream for wired transmission. We adopt the
hyperprior framework in Fig.~\ref{Fig:NIC}, where a hyperprior
$\boldsymbol{z}$ is extracted from features to estimate
their probability distribution for entropy coding.
The coding network is optimized using the R-D objective
in Eq.~\eqref{Eq:RD}, with the Lagrangian multiplier $\lambda$
controlling the trade-off between rate
and reconstruction distortion.

To satisfy varying wired rate constraints, the coding network
must support flexible rate adaptation. Gain-based methods
scale features and hyperpriors using a finite set of gain
parameters, each trained for a specific R-D trade-off.
Consequently, they provide discrete operating points and
cannot directly achieve arbitrary target rates.
To enable continuous rate adaptation, we instead condition
the coding network on the Lagrangian multiplier $\lambda$,
which controls the R-D trade-off in the training objective.
By mapping $\lambda$ to feature modulation parameters,
the network can continuously adjust its intermediate
representations as $\lambda$ varies.

Specifically, for an intermediate feature
$\boldsymbol{l}_{\text{in}}
\in\mathbb{R}^{c_l\times h_l\times w_l}$,
an MLP maps $\lambda$ to channel-wise scaling and bias
parameters:
\begin{equation}
\begin{aligned}
    (\boldsymbol{\alpha},\boldsymbol{\beta})
        &=\operatorname{MLP}(\lambda),\\
    \boldsymbol{l}_{\text{out}}
        &=\boldsymbol{\alpha}\odot
          \boldsymbol{l}_{\text{in}}+\boldsymbol{\beta},
\end{aligned}
\end{equation}
where $\boldsymbol{\alpha},\boldsymbol{\beta}
\in\mathbb{R}^{c_l\times1\times1}$,
and $\odot$ denotes channel-wise multiplication.
The MLP consists of two linear layers with a GELU activation.
As illustrated in Fig.~\ref{Fig:FA_FE}\subref{Fig:FA_FE_a}, each coding block combines
a residual block with an attention layer conditioned on
$\lambda$, allowing the intermediate features to be adjusted
according to the coding preference.

The feature adaptor first
transforms the aligned features $\y$ into
$\boldsymbol{y}_{\lambda}$ according to
Eq.~\eqref{Eq:transformation}. The feature encoder then
extracts the hyperprior $\boldsymbol{z}$ from
$\boldsymbol{y}_{\lambda}$ and uses it to estimate the
distribution parameters for entropy coding.
As shown in Figs.~5(b) and (c), both feature adaptation
and entropy modeling are conditioned on $\lambda$,
so that the representation and its probability model
can adapt together to the selected R-D trade-off.
The relaxed entropy model in Eq.~\eqref{Eq:py} is therefore
extended as
\begin{equation}
    p(\tilde{\boldsymbol{y}}\mid
      \tilde{\boldsymbol{z}},\lambda)
    =
    \prod_i
    \left(
        \mathcal{N}(\tilde{\mu}_i,\tilde{\sigma}_i^2)
        *\mathcal{U}\left(-\frac12,\frac12\right)
    \right)(\tilde{y}_i),
\end{equation}
where
$[\tilde{\boldsymbol{\mu}},\tilde{\boldsymbol{\sigma}}]
=h_s(\tilde{\boldsymbol{z}},\lambda)$.
Here, $\tilde{\boldsymbol{y}}$ is the noisy proxy of
$\boldsymbol{y}_{\lambda}$ used during training.

To train the network across different R-D trade-offs,
we randomly sample $\lambda$ from
$[\lambda_{\text{min}},\lambda_{\text{max}}]$ for each training sample.
The sampled value is used consistently in the R-D objective
and as the conditioning input to the network, enabling
a single model to learn variable-rate coding over the
trained range. At inference, changing $\lambda$ adjusts
the coding rate without switching between separately
trained models. Since encoding and decoding must use
the same conditioning value, $\lambda$ is transmitted
with the bitstream.

\begin{remark}\emph{(Advantages of Using the Lagrangian Multiplier as a Conditioning Input):} First, $\lambda$ directly controls the R-D trade-off in the
optimization objective. Conditioning the network on $\lambda$
therefore allows intermediate features to adapt to the selected
trade-off, helping narrow the performance gap between
variable-rate and single-rate models. Second, sampling
$\lambda$ over a continuous range during training enables
\emph{continuous variable-rate coding} across a broad range
of R-D trade-offs. In contrast, the gain-based schemes use a finite set of gain units trained
at discrete operating points. Finally, this conditioning
mechanism facilitates \emph{fine-grained rate control} by predicting
an appropriate $\lambda$ for the input features and target rate,
as detailed in \Cref{Section:Rate_Control}.
\label{remark:advantages}
\end{remark}

\subsection{Lagrangian Multiplier-Based Rate Control}
\label{Section:Rate_Control}
In this section, we propose to predict an appropriate Lagrangian multiplier $\lambda$ for a given input by a neural network to make the actual coding rate approach the target rate, thus handling wired transmission rate constraints.

\begin{figure}[!t]
\centering
\includegraphics[width=0.45\textwidth]{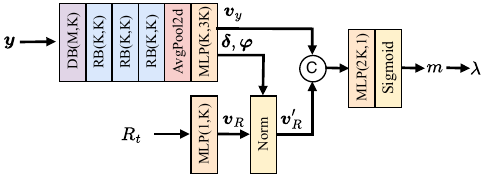}
\caption{Architecture of the Lagrangian multiplier predictor
($K=128$).}
\label{Fig:HP}
\end{figure}

Although varying $\lambda$ enables continuous rate adaptation,
selecting an appropriate value to meet a given wired rate
constraint remains challenging. A value that is too small may
underutilize the available bandwidth, whereas a value that is
too large may produce a bitstream exceeding the rate budget.
However, finding a suitable $\lambda$ through repeated encoding
incurs significant additional computation and latency, which motivates us to predict an appropriate $\lambda$ before encoding rather than searching for it through repeated encoding.

Since $\lambda$ controls the R-D trade-off, larger values
generally yield higher coding rates. Previous work
\cite{9998500} models this relationship as
\begin{equation}
    \lambda=\kappa\left(e^{\epsilon R}-1\right),
    \label{Eq:relation}
\end{equation}
where $\kappa$ and $\epsilon$ are fitted parameters.
However, the coding rate also depends on the input content. The same $\lambda$ may yield different rates for images with
different complexity and texture characteristics. Therefore,
predicting $\lambda$ requires considering both the target rate
and the features to be encoded.

To capture this content-dependent relationship, we introduce
a Lagrangian multiplier predictor $\operatorname{LMP}(\cdot)$,
which takes the target rate $R_t$ and aligned features $\y$
as inputs and predicts $\lambda$ as formulated in
Eq.~\eqref{Eq:prediction}. The predicted $\lambda$ then
conditions the feature adaptor and feature encoder to produce
a bitstream whose rate approximates $R_t$.
Motivated by the exponential relationship in
Eq.~\eqref{Eq:relation}, we use $\lambda$ in logarithmic space as the
conditioning input to all $\lambda$-conditioned modules.

As illustrated in Fig.~\ref{Fig:HP}, the Lagrangian multiplier
predictor takes the aligned features $\y$ and target rate $R_t$
as inputs. The features $\y$ are processed by residual blocks,
followed by average pooling and an MLP to obtain a feature
vector ${\vect}_{y}\in\mathbb{R}^K$. In parallel, another MLP
maps $R_t$ to a rate vector ${\vect}_{R}\in\mathbb{R}^K$.
To account for the content dependence of the coding rate,
the predictor also derives normalization parameters
$\boldsymbol{\delta},\boldsymbol{\varphi}\in\mathbb{R}^K$
from $\y$ and applies them to the rate vector:
\begin{equation}
    {\vect}^{\prime}_{R}
    =\frac{{\vect}_{R}-\boldsymbol{\delta}}
           {\boldsymbol{\varphi}},
\end{equation}
where division is performed element-wise.
The normalized rate vector ${\vect}^{\prime}_{R}$ is then
concatenated with ${\vect}_{y}$ to produce a scalar score
$m\in[0,1]$. This score is mapped to the trained range
$[\lambda_{\text{min}},\lambda_{\text{max}}]$ to obtain the predicted
Lagrangian multiplier:
\begin{equation}
    \lambda
    =(\lambda_{\text{max}}-\lambda_{\text{min}})m +\lambda_{\text{min}}.
\end{equation}

\begin{figure}[t]
\centering
\subfloat[$\lambda$ sampling.]{
\centering
\includegraphics[width=0.27\linewidth]{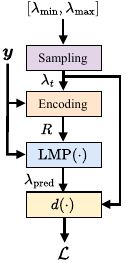}%
\label{Fig:Sampling_a}}
\hspace{1cm}
\subfloat[Rate sampling.]{
\centering
\includegraphics[width=0.275\linewidth]{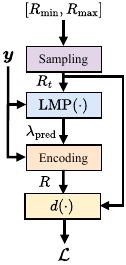}%
\label{Fig:Sampling_b}}
\caption{Two different strategies for training the Lagrangian multiplier predictor.}
\label{Fig:Sampling}
\end{figure}

The Lagrangian multiplier predictor is trained after the
variable-rate coding model, with all other network parameters
fixed. As illustrated in Fig.~\ref{Fig:Sampling}, two training
strategies are considered, differing in the sampled target
and training objective.

\subsubsection{$\lambda$ Sampling}
A target multiplier $\lambda_t$ is sampled from
$[\lambda_{\text{min}},\lambda_{\text{max}}]$ and used to encode
the aligned features $\y$, yielding a coding rate $R$.
The predictor then takes $\y$ and $R$ as inputs to estimate
$\lambda_{\text{pred}}$. The training objective measures
the discrepancy between the sampled and predicted multipliers:
\begin{equation}
    \mathcal{L}=d(\lambda_t,\lambda_{\text{pred}}).
    \label{Eq:lambda_bias}
\end{equation}

\subsubsection{Rate Sampling}
A target rate $R_t$ is sampled from
$[R_{\text{min}},R_{\text{max}}]$ and supplied to the predictor
together with $\y$ to obtain $\lambda_{\text{pred}}$.
The coding model then encodes $\y$ conditioned on
$\lambda_{\text{pred}}$, yielding a rate $R$.
The training objective directly measures the discrepancy
between the target and resulting rates:
\begin{equation}
    \mathcal{L}=d(R_t,R).
    \label{Eq:R_bias}
\end{equation}

We adopt rate sampling because its objective directly
reflects the desired rate control accuracy.
Specifically, we use the normalized squared error:
\begin{equation}
    \mathcal{L}=
    \left(\frac{R-R_t}{R_t}\right)^2.
    \label{Eq:normalization}
\end{equation}
This loss penalizes relative rate deviations, assigning
greater weight to the same absolute error at smaller
target rates. Once trained, the predictor selects
$\lambda$ from the aligned features $\y$ and target rate
$R_t$ to bring the coding rate close to the target.

\begin{remark}\emph{(Rate Control via Lagrangian Multiplier Prediction):}
We use a neural network to predict an appropriate Lagrangian
multiplier $\lambda$ from the aligned features $\y$ and
target rate $R_t$, capturing the content-dependent
$R$--$\lambda$ relationship. The predicted $\lambda$ then
conditions the feature adaptation and coding modules to
produce a bitstream whose rate closely matches $R_t$.
\end{remark}

\begin{figure*}[t]
\centering
\subfloat[Diagram of $\lambda$-conditional context model.]{\includegraphics[width=0.23\linewidth]{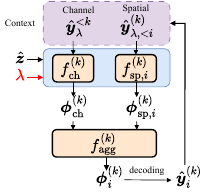}%
\label{Fig:context_a}}
\hspace{0.1cm}
\subfloat[Channel context model.]{\includegraphics[width=0.25\linewidth]{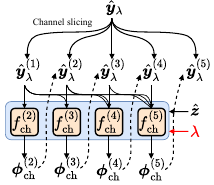}%
\label{Fig:context_b}}
\hspace{0.1cm}
\subfloat[Spatial context model.]{\includegraphics[width=0.23\linewidth]{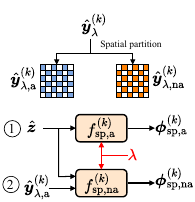}%
\label{Fig:context_c}}
\caption{Context modeling for high-resolution images.}
\label{Fig:context}
\end{figure*}

\subsection{Extension to High-Resolution Image Coding}
For high-resolution images, the dependencies among pixels become more prominent, making it difficult to achieve effective coding by modeling latent feature distribution with only hyperprior $\z$. Based on ELIC\cite{ELIC}, we improve our feature encoder and decoder for better coding of high-resolution images by fully exploiting contexts, i.e., previously decoded information while maintaining their continuous variable-rate coding function.

To encode the $i$-th element $y_{\lambda,i}^{(k)}$ in the $k$-th channel chunk, we use previously decoded channel chunks $\yhat_{\lambda}^{<k}$ and spatial
elements $\yhat_{\lambda,<i}^{(k)}$ as contexts. The context model utilizes functions $f_\text{ch}^{(k)}(\cdot)$ and $f_{\text{sp},i}^{(k)}(\cdot)$ to process channel context $\yhat^{<k}_{\lambda}$ and spatial context $\yhat^{(k)}_{\lambda,<i}$ respectively, both of which are parameterized by neural networks:
\begin{align}
\begin{aligned}
&\operatorname{Channel}:\boldsymbol{\phi}^{(k)}_\text{ch}=f_\text{ch}^{(k)}(\yhat^{<k}_{\lambda},\zhat,\lambda), \\
&\operatorname{Spatial}:\boldsymbol{\phi}^{(k)}_{\text{sp},i}=f_{\text{sp},i}^{(k)}(\yhat^{(k)}_{\lambda,<i},\zhat,\lambda),
\end{aligned}
\end{align}
where $\boldsymbol{\phi}^{(k)}_\text{ch}$ and $\boldsymbol{\phi}^{(k)}_{\text{sp},i}$ are entropy parameters estimated by considering channel contexts and spatial contexts, respectively. Note that since the hyperprior $\zhat$ carries global information of $\yl$, it is supplied to both networks to improve entropy estimation. To facilitate variable-rate coding, $\lambda$ is also injected into both networks by adding several coding blocks illustrated in Fig.~\ref{Fig:FA_FE}\subref{Fig:FA_FE_a}, allowing the entropy models to adapt to different R-D trade-offs. Then, $\boldsymbol{\phi}^{(k)}_\text{ch}$ and $\boldsymbol{\phi}^{(k)}_{\text{sp},i}$ are aggregated by another network $f^{(k)}_\text{agg}(\cdot)$ to produce the final entropy parameters $\boldsymbol{\phi}^{(k)}_i$ for entropy coding of $\hat{y}_{\lambda,i}^{(k)}$:
\begin{equation}
  \boldsymbol{\phi}^{(k)}_i=f^{(k)}_\text{agg}(\boldsymbol{\phi}^{(k)}_\text{ch},\boldsymbol{\phi}^{(k)}_{\text{sp},i}).
\end{equation}
Once $\hat{y}_{\lambda,i}^{(k)}$ is decoded, it provides additional context for the entropy encoding and decoding of subsequent features. The operational diagram is shown in Fig.~\ref{Fig:context}\subref{Fig:context_a}.

Fig.~\ref{Fig:context}\subref{Fig:context_b} illustrates the chunk-level utilization of channel contexts. Specifically, $\yhat_\lambda$ is divided into 5 chunks $(\yhat_\lambda^{(1)},\yhat_\lambda^{(2)},\yhat_\lambda^{(3)},\yhat_\lambda^{(4)},\yhat_\lambda^{(5)})$. The entropy parameters $\boldsymbol{\phi}_\text{ch}^{(i)}$ of the $i$-th chunk are generated by feeding all previously decoded chunks $(\yhat_\lambda^{(1)},...,\yhat_\lambda^{(i-1)})$, together with $\zhat$ and $\lambda$, into the context model in a single forward pass. These parameters are then used for the coding of the current chunk $\yhat_\lambda^{(i)}$.

Spatial context is modeled using a checkerboard partition, as illustrated in Fig.~\ref{Fig:context}\subref{Fig:context_c}. Given the $k$-th chunk $\yhat_\lambda^{(k)}$, anchor features $\yhat_{\lambda,\text{a}}^{(k)}$ and non-anchor features $\yhat_{\lambda,\text{na}}^{(k)}$ are obtained using a checkerboard spatial partition. The entropy parameters of the anchor features $\boldsymbol{\phi}_\text{sp,a}^{(k)}$ are produced by the network $f_{\text{sp,a}}^{(k)}(\cdot)$ with input $\zhat$ and $\lambda$. The entropy parameters of the non-anchor features $\boldsymbol{\phi}_\text{sp,na}^{(k)}$ are produced by the network $f_{\text{sp,na}}^{(k)}(\cdot)$ with input $\zhat$, $\lambda$ and anchor features $\yhat_{\lambda,\text{a}}^{(k)}$:
\begin{align}
\begin{aligned}
&\operatorname{Anchor}:\boldsymbol{\phi}_\text{sp,a}^{(k)}=f_{\text{sp,a}}^{(k)}(\emptyset,\zhat,\lambda), \\
&\operatorname{Non-Anchor}:\boldsymbol{\phi}_\text{sp,na}^{(k)}=f_{\text{sp,na}}^{(k)}(\yhat_{\lambda,\text{a}}^{(k)},\zhat,\lambda).
\end{aligned}
\end{align}

\subsection{Training Strategy}
In this section, we present a three-stage training procedure
for DARA-DJSCC. We first train wireless transmission with
distribution alignment, then introduce continuous variable-rate
feature coding, and finally train the rate controller.
Each stage retains the parameters learned in the preceding
stage to facilitate stable optimization.

\subsubsection{Fully Wireless JSCC Coding}
We first train the DJSCC modules and DAT without wired
feature coding. A fixed pre-trained NIC feature extractor
provides reference features $\y_{\text{sup}}$ to supervise
distribution alignment. The transmission pipeline is
\begin{equation}
    \x\to\s
    \xrightarrow{W_1(\cdot,\eta_1)}\shat
    \to\y\to\boldsymbol{u}
    \xrightarrow{W_2(\cdot,\eta_2)}\hat{\boldsymbol{u}}
    \to\tilde{\x},
\end{equation}
where $W_1$ and $W_2$ denote the source and destination
wireless access channels with SNRs $\eta_1$ and $\eta_2$,
respectively. We optimize $\boldsymbol{\theta}$ and
$\boldsymbol{\tau}$, with both SNRs randomly sampled from
$[\eta_{\text{min}},\eta_{\text{max}}]$.
The training loss combines reconstruction distortion
and the alignment loss in Eq.~\eqref{eq:align_loss}:
\begin{equation}
    \mathcal{L}_1
    =\pi\mathcal{L}_{\text{MSE}}(\x,\tilde{\x})
    +(1-\pi)\mathcal{L}_{\text{align}}
    (\y,\y_{\text{sup}}),
\end{equation}
where $\pi=0.67$ in our experiments.

\subsubsection{Continuous Variable-Rate Coding}
Building on the aligned wireless representations, we introduce
the feature adaptor and feature encoder-decoder pair to support
wired transmission:
\begin{equation}
    \x\to\s
    \xrightarrow{W_1(\cdot,\eta_1)}\shat
    \to\y\to\bit
    \xrightarrow{W_c}\bit
    \to\boldsymbol{u}
    \xrightarrow{W_2(\cdot,\eta_2)}\hat{\boldsymbol{u}}
    \to\xhat,
\end{equation}
where $W_c$ denotes reliable wired forwarding through
the core network. We jointly optimize
$\boldsymbol{\theta}$, $\boldsymbol{\tau}$, and
$\boldsymbol{\omega}$, while leaving LMP for the final stage.
The wireless SNRs are sampled as in the first stage,
and $\lambda$ is sampled uniformly in logarithmic space
over $[\lambda_{\text{min}},\lambda_{\text{max}}]$.
The loss is the normalized R-D objective:
\begin{equation}
    \mathcal{L}_2
    =\frac{\lambda}{1+\lambda}d(\x,\xhat)
    +\frac{1}{1+\lambda}
    R(\bit_{\yhatl},\bit_{\zhat}),
\end{equation}
where $\bit_{\yhatl}$ and $\bit_{\zhat}$ denote the
bitstreams associated with the quantized features
$\yhatl$ and hyperprior $\zhat$, respectively.

\subsubsection{Rate Control}
Finally, we freeze the trained transmission and coding modules
and optimize only the LMP parameters $\boldsymbol{\xi}$.
Given the aligned features $\y$ and a target wired rate $R_t$,
LMP predicts $\lambda_{\text{pred}}$, which conditions
the coding network. The wireless SNRs are sampled as before,
and $R_t$ is sampled from $[R_{\text{min}},R_{\text{max}}]$.
We minimize the normalized rate error in
Eq.~\eqref{Eq:normalization} to bring the resulting coding
rate close to $R_t$.

\section{Numerical Results and Analysis} 
In this section, we first detail the experimental setup, then present and analyze the quantitative comparison results over advanced neural network-based approaches and traditional baselines, demonstrating effectiveness of proposed DARA for transmission in wired links. 
\label{Section:numerical_results}
\subsection{Experimental Settings}
\subsubsection{Low-Resolution Images} We use CIFAR-10 as the main dataset for training and validation, which consists of 50,000 training samples and 10,000 testing samples with resolution $32\times32$. Since our module is compatible with most mainstream DJSCC methods, we adopt the model proposed in \cite{Bian_Hybird} as the DJSCC encoder and decoder to perform wireless transmission. The channel-input vectors $\s$ and $\boldsymbol{u}$ have the same size of $8\times8$ with $24$ channels, which corresponds to $768$ complex-valued transmission symbols and wireless CBR of $\rho_1=\rho_2=0.25$. The sizes of recovered feature $\y$, adapted feature $\yl$ and hyperprior $\z$ are $8\times8$, $8\times8$ and $4\times4$, respectively. Both $\y$ and $\y_\lambda$ have $M=256$ channels,
while $\boldsymbol{z}$ has $N=192$ channels. Unless otherwise specified, the main experiments are conducted on this dataset.

\subsubsection{High-Resolution Images} We use COCO\cite{COCO} as the training dataset and CLIC Professional Validation \cite{clic2020} with approximately 2K resolution as the validation dataset. COCO consists of about 118,000 images with various resolutions and we randomly crop these images with the size of $256\times256$ for training. We adopt bandwidth-adaptive DJSCC\cite{NTSCC}, i.e., NTSCC as the DJSCC encoder and decoder to overcome the performance saturation problem for wireless transmission. The number of wireless transmission symbols varies according to the complexity of images and we just set the average wireless CBR as $\rho_1=\rho_2=0.08$. $\y$ and $\yl$ have a spatial resolution that is downsampled by a factor of 16 relative to the original image, while $\z$ is downsampled by a factor of 32. Both $\y$ and $\yl$ have $M=256$ channels, while $\z$ has $N=192$ channels.

\begin{figure*}[t]
\centering
\subfloat[$\eta_1=\eta_2=2$ dB.]{\includegraphics[width=0.34\linewidth]{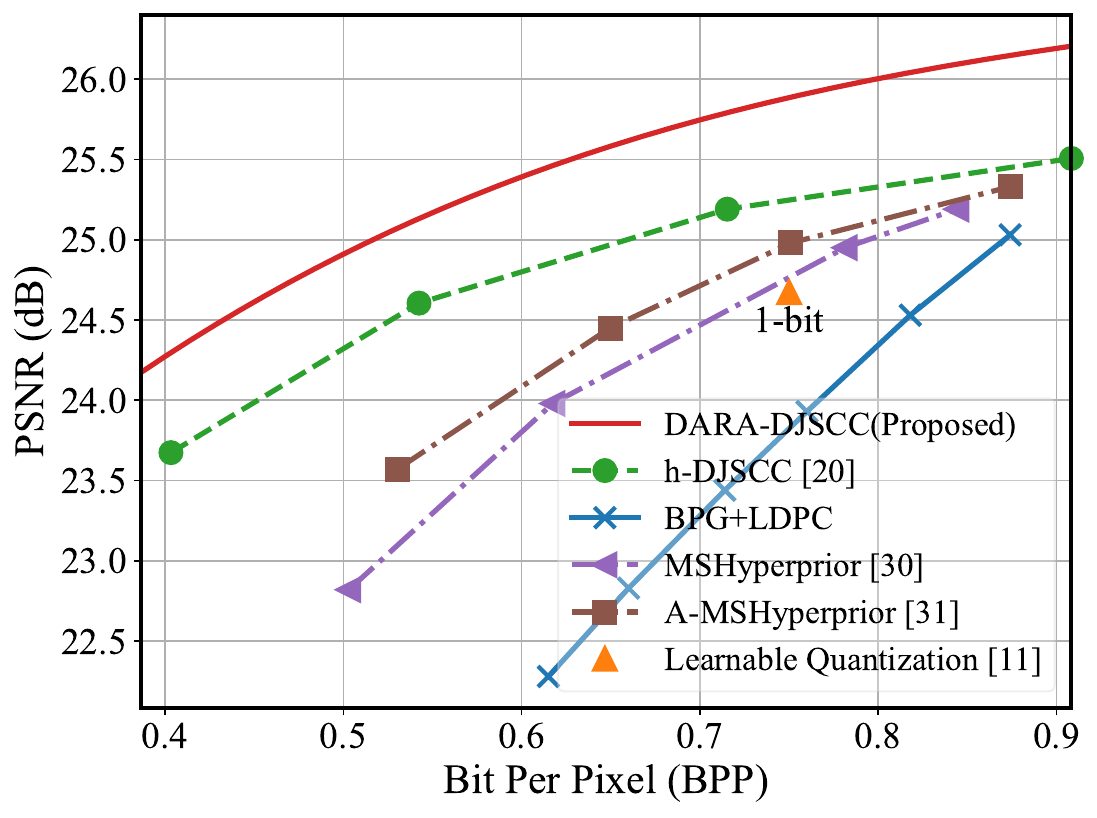}%
\label{Fig:T_2dB_cifar}}
\hfil
\subfloat[$\eta_1=\eta_2=5$ dB.]{\includegraphics[width=0.33\linewidth]{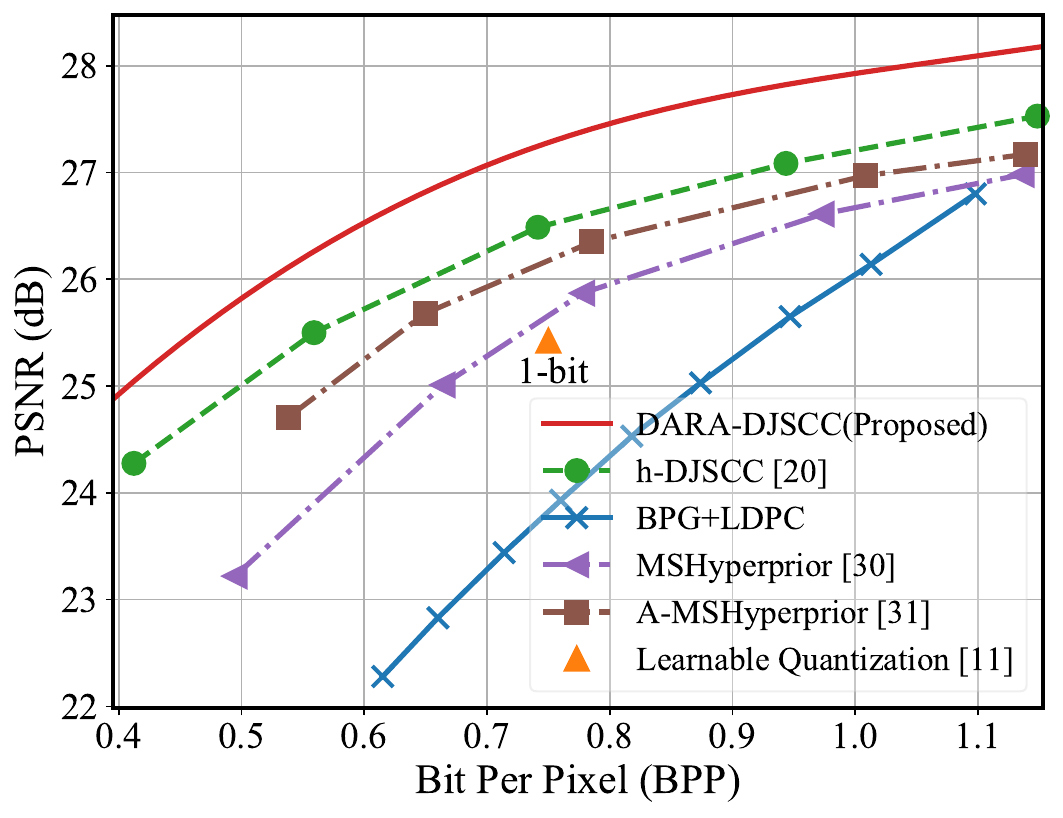}%
\label{Fig:T_5dB_cifar}}
\hfill
\subfloat[$\eta_1=\eta_2=10$ dB.]{\includegraphics[width=0.33\linewidth]{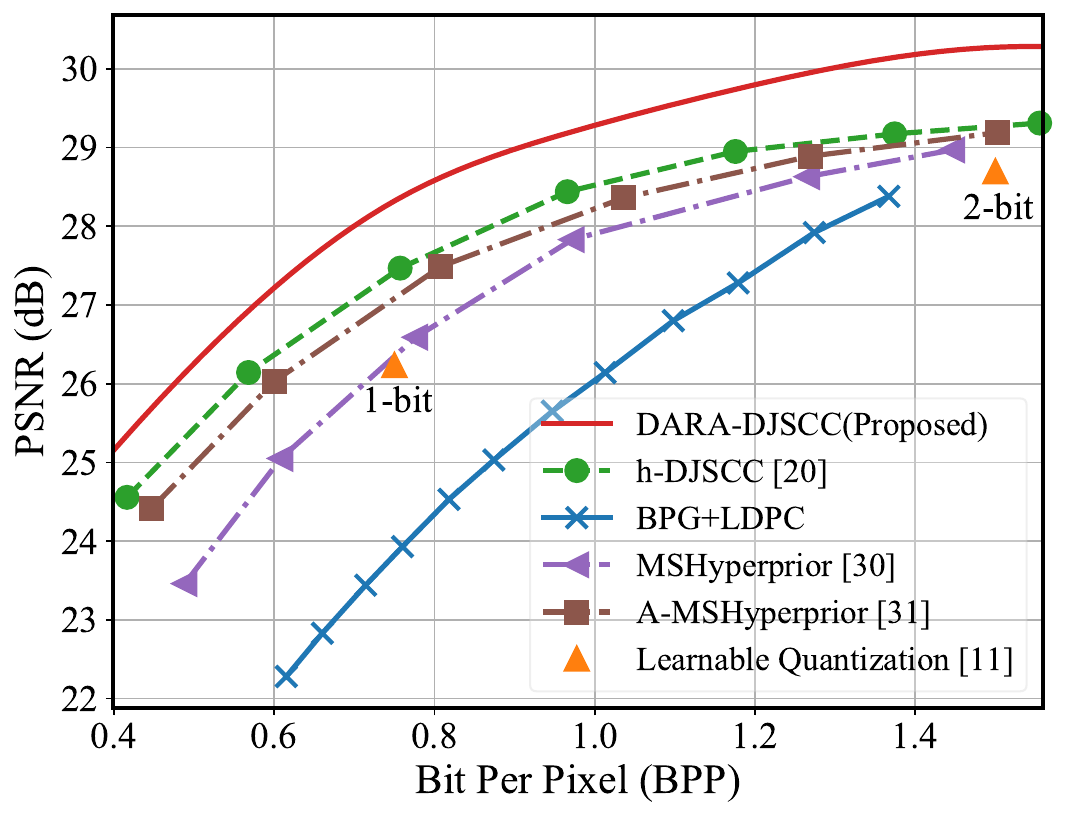}%
\label{Fig:T_10dB_cifar}}
\caption{R-D performance comparison of different transmission schemes under specific SNRs over AWGN channels on CIFAR-10 dataset.}
\label{Fig:T_bpp_cifar}
\end{figure*}

\subsubsection{Training Details} For all training stages, the Adam optimizer is employed with an initial learning rate of $1\times{10^{-4}}$. If the loss does not decrease for two consecutive epochs, the learning rate is decayed by a factor of 0.5. We stop the optimization of each stage when the learning rate reaches $1\times{10^{-6}}$. A batch size of 32 is used for low-resolution images, and 16 for high-resolution images. During variable-rate coding training, $\lambda$ is sampled
uniformly in logarithmic space over the range from 16 to 4096.

\subsubsection{Constraint Variables and Evaluation Metrics}
Due to variations in image resolutions, we use average transmission resource cost for each pixel to evaluate transmission performance. Thus, wireless constraints and wired constraints can be enforced by limiting CBR and BPP, respectively. Higher CBR and BPP values imply greater consumption of wireless and wired resources, and vice versa. PSNR is used to evaluate reconstruction quality, which can be computed by:
\begin{equation}
    \operatorname{PSNR}(\x,\xhat)
    =10\log_{10}
    \left(\frac{255^2}{\operatorname{MSE}(\x,\xhat)}\right).
\end{equation}
where $\operatorname{MSE}(\x,\xhat)$ denotes MSE between the original image $\x$ and the reconstructed image $\xhat$.

\subsection{Hybrid Transmission Performance Comparison}
In this subsection, we evaluate the hybrid transmission performance of DARA-DJSCC against various baselines in wireless-wired network scenarios. We analyze the R-D performance under specific wireless channel conditions and further investigate the robustness of different transmission schemes to varying wireless SNRs.

The comparison baselines include:
\subsubsection{h-DJSCC} An advanced hybrid DJSCC transmission scheme proposed in \cite{Bian_JSAC}, which is optimized for mobile multi-hop networks.
\subsubsection{BPG+LDPC} The traditional digital transmission scheme. BPG and low-density parity-check (LDPC) are used for source and channel coding, respectively. For the core network, channel decoding is performed first and only information bits are transmitted through wired links.
\subsubsection{Learnable Quantization} A digital DJSCC transmission scheme, where a learnable quantization \cite{digital-sc} is introduced to quantize analog DJSCC symbols in bits for transmission in the core network. At the source access node, received analog DJSCC symbols are directly processed by fixed quantization, i.e., each symbol is quantized into a fixed number of bits (the quantization values can be learned), and then transmitted through wired links.

\subsubsection{Modular Transmission} The modular transmission scheme combining JSCC with classical NIC models, including MeanScaleHyperprior\cite{MSHyperPrior} (denoted as MSHyperprior), Attention-based MeanScaleHyperprior \cite{Cheng2020Attention} (denoted as A-MSHyperprior) and ELIC\cite{ELIC}. A JSCC model is utilized for wireless transmission, while various NIC models are used independently for wired segments. In this scheme, the JSCC and NIC modules are optimized for their respective objectives without any joint training or distribution alignment.

\begin{figure*}[!t]
\centering
\begin{minipage}[t]{0.65\textwidth}
    \vspace{0pt}
    \centering
    \subfloat[$\eta_2=10$dB and $\eta_1$ varies.]{%
        \begin{minipage}{0.5\linewidth}
            \centering
            \includegraphics[width=\linewidth]{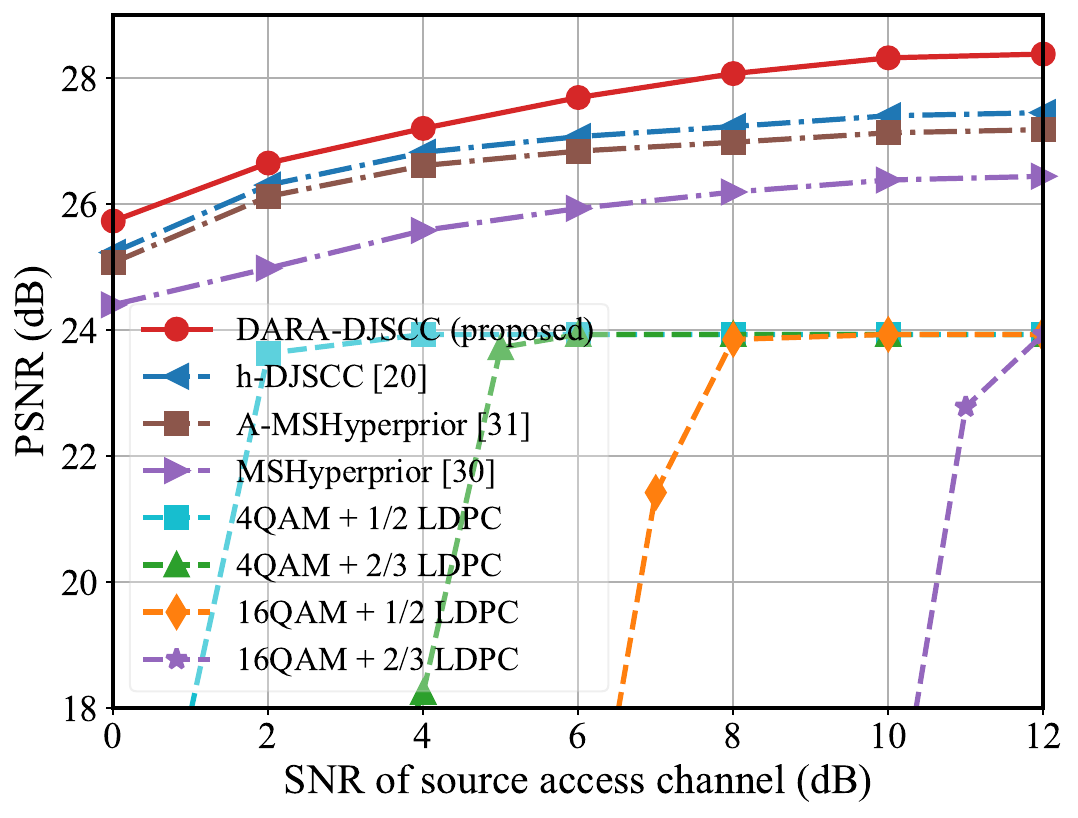}
        \end{minipage}%
        \label{Fig:SNR1_cifar}%
    }%
    \hfill
    \subfloat[$\eta_1=10$dB and $\eta_2$ varies.]{%
        \begin{minipage}{0.5\linewidth}
            \centering
            \includegraphics[width=\linewidth]{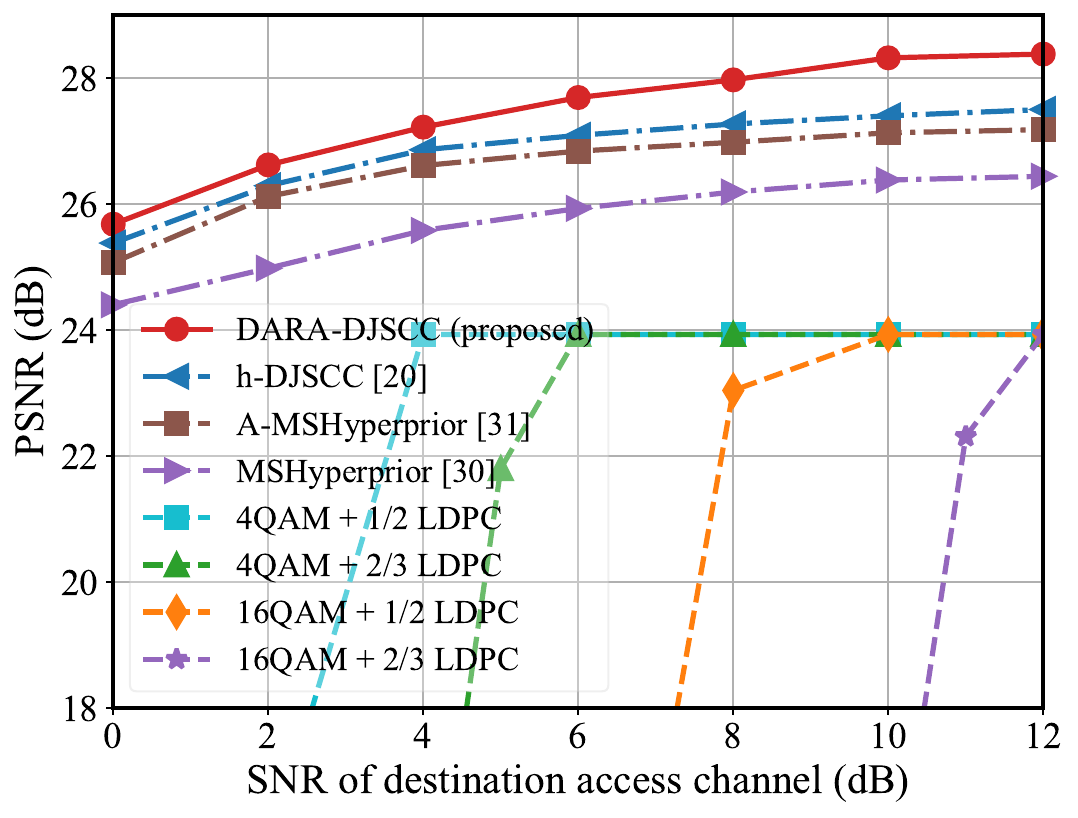}
        \end{minipage}%
        \label{Fig:SNR2_cifar}%
    }
    \caption{Hybrid transmission performance on CIFAR-10
    under varying wireless SNRs, with the wired rate fixed
    at 0.75~BPP.}
    \label{Fig:SNR_cifar}
\end{minipage}%
\hfill
\begin{minipage}[t]{0.33\textwidth}
    \vspace{5pt}
    \centering
    \vspace{6pt}
    \includegraphics[width=\linewidth]{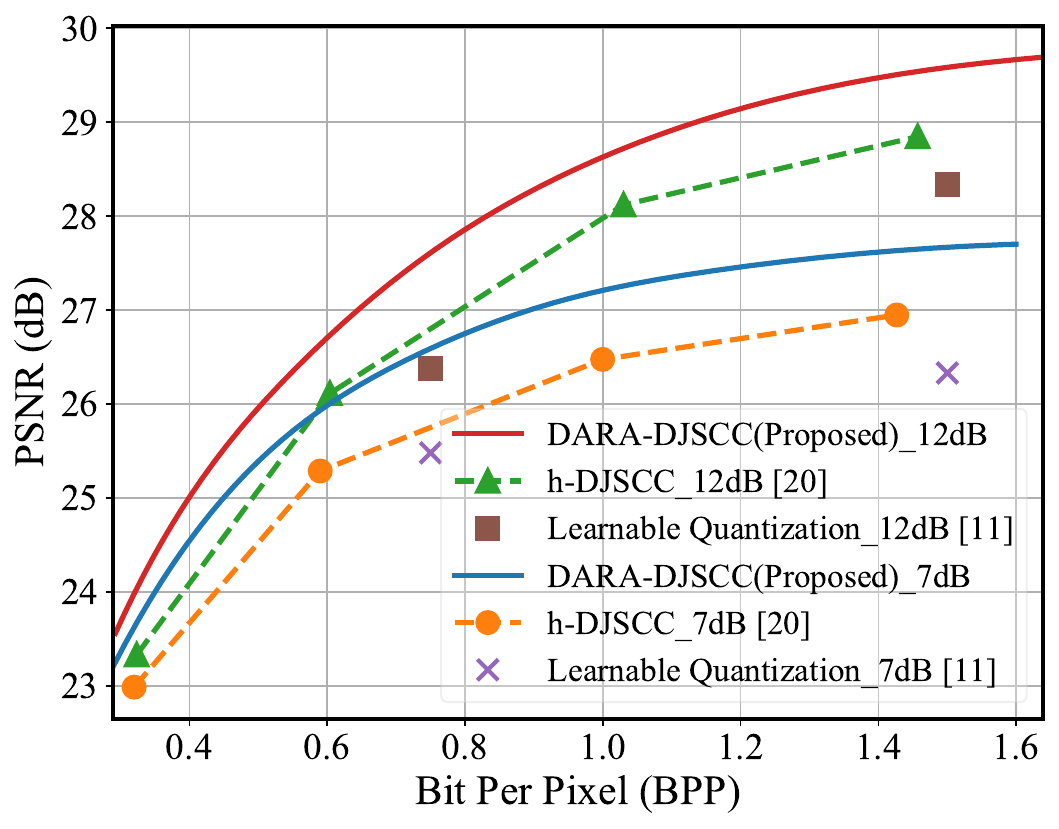}
    \caption{Hybrid transmission performance over Rayleigh
    fading channels on the CIFAR-10 dataset.}
    \label{Fig:Fading}
\end{minipage}
\end{figure*}

Except for modular baselines, whose JSCC and NIC components
are trained separately, all neural transmission schemes are
jointly trained E2E. The JSCC components are trained
under mixed SNRs ranging from $0$dB to $10$dB.

We first verify R-D performance of different methods under fixed wireless conditions over AWGN channels, which is illustrated in Fig. \ref{Fig:T_bpp_cifar}. For simplicity, the source access wireless channel and destination access wireless channel are set to the identical condition, i.e., $\eta_1=\eta_2=2,5,10\text{dB}$. BPP is used to measure bit cost as well as transmission rate in wired links.

Across all tested SNR values, DARA-DJSCC consistently achieves superior R-D performance compared to baselines, with PSNR generally increasing with higher BPP. This advantage stems from DARA directly embedding Lagrangian multiplier $\lambda$ into the coding process via an attention-based mechanism, yielding sufficient nonlinearity and better coding performance, especially noticeable at higher BPPs. The modular transmission schemes, which independently employ JSCC for wireless segments and classical NIC models for wired links, also underperform both DARA-DJSCC and h-DJSCC due to the lack of joint optimization.  Furthermore, h-DJSCC relies on discrete gain units, and learnable quantization uses fixed quantization levels (e.g., 1-bit, 2-bit), limiting them to a few specific rates. In contrast, because of the continuous range of $\lambda$, DARA enables continuous rate coding, thus supporting continuous rate adaptation within the trained range. The traditional digital scheme exhibits rapid performance deterioration, particularly in the low BPP range, resulting in a substantial performance gap relative to neural network-based schemes.

Then, we investigate the hybrid transmission performance under varying wireless channel conditions with fixed BPP approximately to 0.75 of wired segments, as illustrated in Fig. \ref{Fig:SNR_cifar}. Here, we evaluate the robustness of different schemes by fixing the SNR of one wireless channel ($\eta_1$ or $\eta_2$) to 10dB while varying the other. (288, 576) and (432, 576) LDPC codes are employed. Generally, DARA-DJSCC outperforms other schemes across all varying SNR ranges, demonstrating its superior robustness and adaptability. Consistent with the R-D analysis, modular transmission baselines also exhibit lower performance than DARA-DJSCC and h-DJSCC, as they lack the joint optimization required to handle the complex distribution changes during the wireless-to-wired transition. It should be noted that traditional digital baselines combining different quadrature amplitude modulation (QAM) orders with LDPC codes of different rates exhibit a sharp ``cliff effect'' where their performance abruptly degrades below a certain SNR threshold. In contrast, DJSCC-based schemes like DARA-DJSCC and h-DJSCC show a graceful performance degradation. Even when compared with the most robust traditional digital baseline (4QAM+1/2LDPC), DARA-DJSCC achieves a PSNR gain of more than 3dB.

Building upon the analysis of fixed and varying AWGN conditions, Fig. \ref{Fig:Fading} further extends our investigation to the more challenging scenario of Rayleigh fading channels. It presents two sets of curves corresponding to $\eta_1=\eta_2=7,12$dB conditions. As depicted in Fig. \ref{Fig:Fading}, DARA-DJSCC demonstrates strong adaptability to transmission over Rayleigh fading channels, achieving the best R-D performance among the three neural network-based methods. The h-DJSCC method performs second-best, generally outperforming learnable quantization across all BPP regions. Notably, learnable quantization exhibits the lowest PSNR values across all BPP points and SNR conditions, highlighting its comparatively limited performance in this challenging environment.

\begin{figure*}[!t]
\centering
\begin{minipage}[t]{0.63\textwidth}
    \vspace{0pt}
    \centering
    \subfloat[$\eta_1=\eta_2=10$dB.]{%
        \begin{minipage}{0.48\linewidth}
            \centering
            \includegraphics[width=\linewidth]{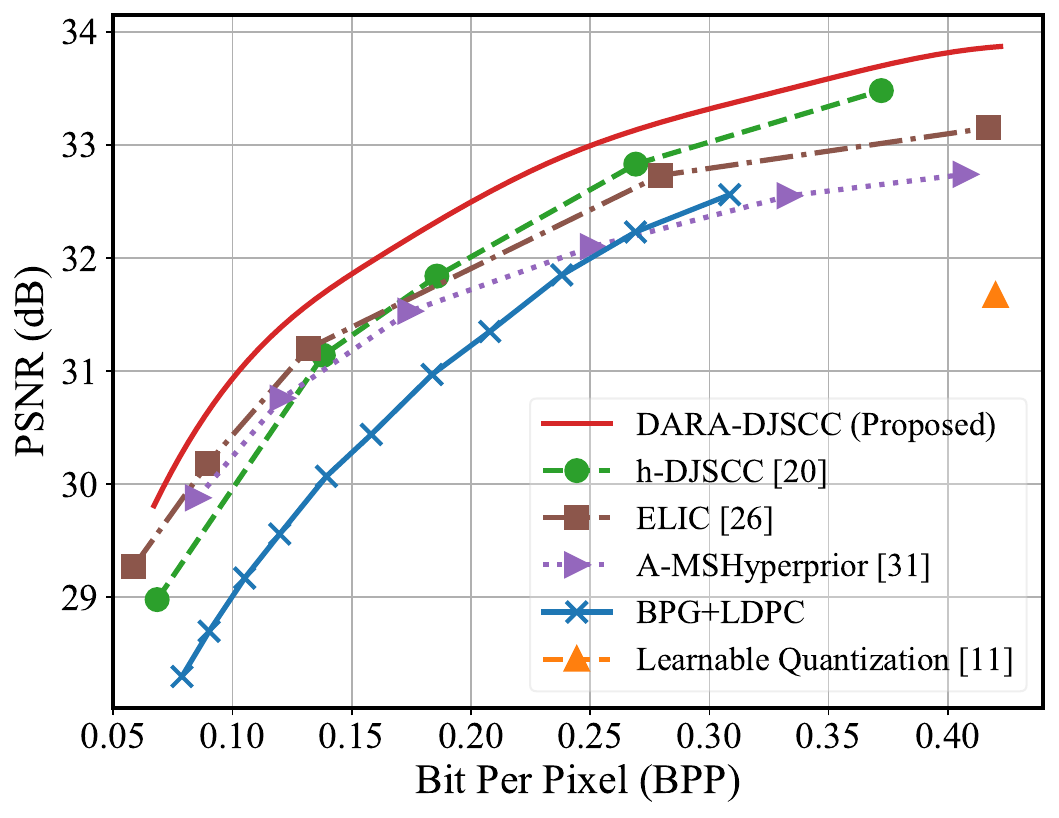}
        \end{minipage}%
        \label{Fig:BPP_Kodak}%
    }%
    \hspace{0.02\linewidth}%
    \subfloat[$\eta_1=10$dB, $\text{BPP}=0.18$.]{%
        \begin{minipage}{0.48\linewidth}
            \centering
            \includegraphics[width=\linewidth]{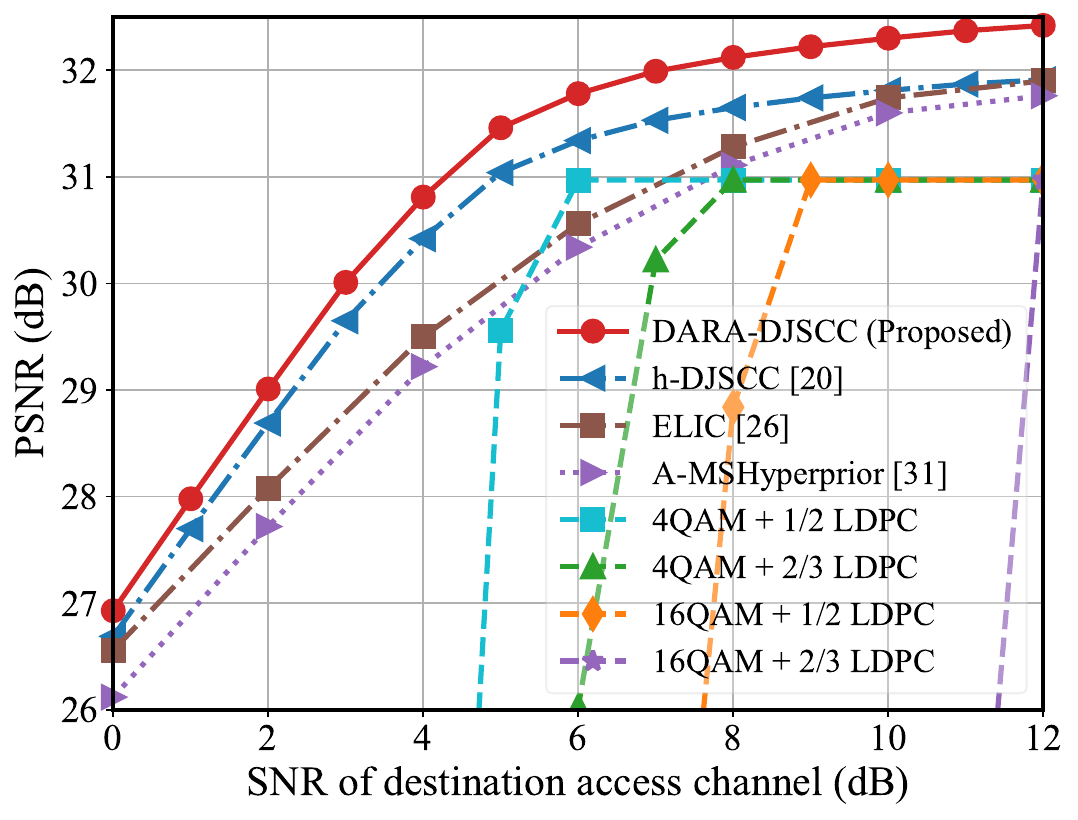}
        \end{minipage}%
        \label{Fig:SNR_Kodak}%
    }
    \caption{Hybrid transmission performance comparison
    on the CLIC dataset.}
    \label{Fig:Kodak}
\end{minipage}%
\hspace{0.02\textwidth}%
\begin{minipage}[t]{0.32\textwidth}
    \vspace{3pt}
    \centering
    \vspace{0pt}
    \includegraphics[width=\linewidth]{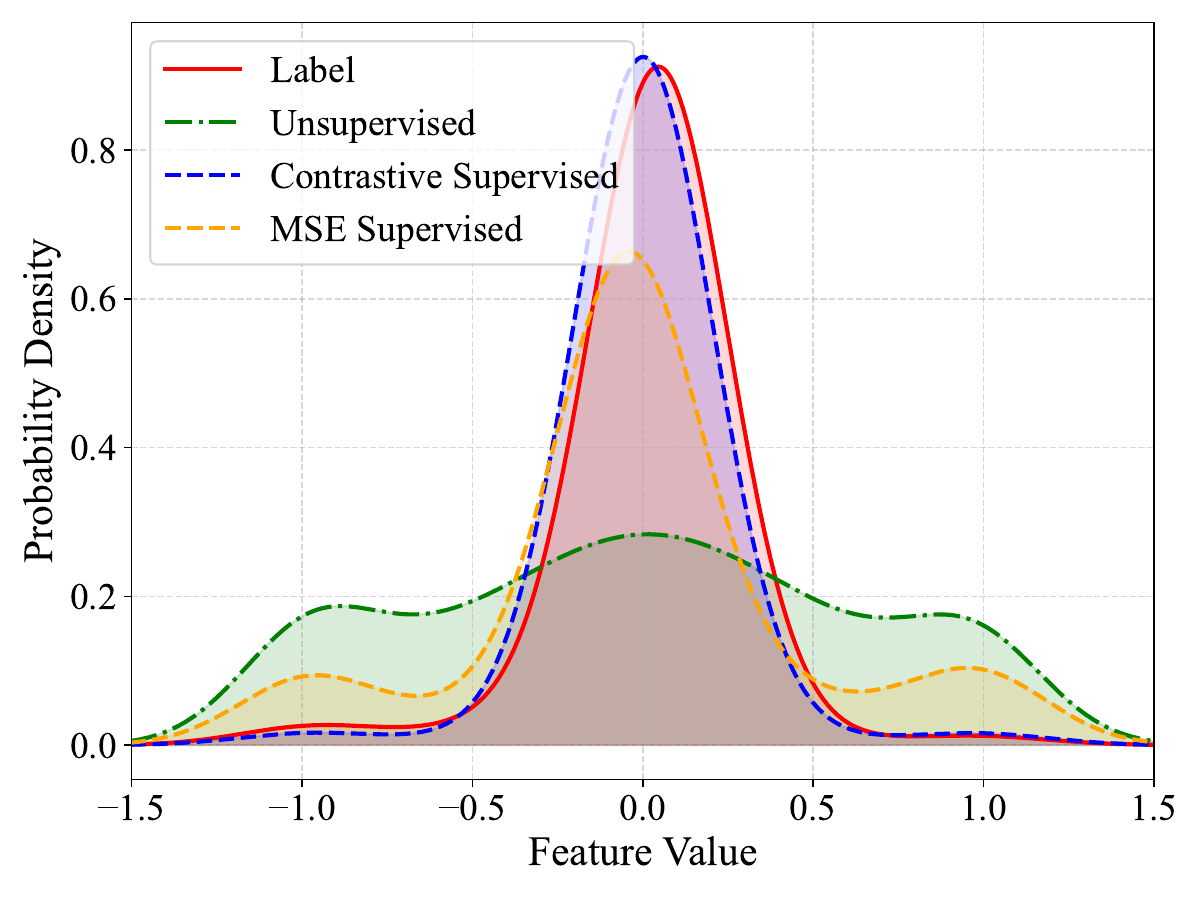}
    \caption{Distributions of aligned features under different
    supervision strategies on CIFAR-10 at an SNR of 10dB.}
    \label{Fig:visualization_ablation}
\end{minipage}
\end{figure*}

\begin{table}[!t]
\centering
\caption{Ablation study of distribution alignment.}
\label{tab:ablation}
\setlength{\tabcolsep}{3.5pt}
\begin{tabular}{cccc}
\toprule
\textbf{SNR (dB)} & \textbf{Unsupervised} & \textbf{MSE Supervised} & \textbf{Contrastive Supervised} \\
\midrule
0  & 25.50 & 25.65 & \textbf{25.86} \\
2  & 26.56 & 26.71 & \textbf{26.91} \\
4  & 27.12 & 27.27 & \textbf{27.54} \\
6  & 27.53 & 27.68 & \textbf{27.93} \\
8  & 27.83 & 27.99 & \textbf{28.28} \\
10 & 27.94 & 28.20 & \textbf{28.57} \\
\bottomrule
\end{tabular}
\end{table}

To verify the effectiveness of the proposed DAT, we evaluate models trained under diverse supervision strategies. In this configuration, the destination wireless access channel is omitted to isolate the impact of the adaptor and simulate scenarios where the receiver resides within the core network. We fix the wired segment BPP at 0.8 and vary the source wireless channel quality to generate DJSCC symbols under diverse conditions. As illustrated in Table \ref{tab:ablation}, supervised alignment guided by a pre-trained neural compressor consistently outperforms unsupervised training. Specifically, among supervised strategies, contrastive learning-based alignment achieves higher PSNR gain over conventional MSE-based supervision. This superiority demonstrates that aligning features at the distribution level is more effective for hybrid adaptation than minimizing element-wise feature differences. The performance gap becomes increasingly pronounced as the channel quality improves. For instance, at a wireless SNR of 10dB, contrastive supervision improves PSNR
by 0.63dB over unsupervised training and 0.37dB over MSE-based supervision.

Furthermore, Fig. \ref{Fig:Kodak} extends our evaluation to high-resolution images using the CLIC dataset, demonstrating the scalability and effectiveness of DARA-DJSCC. Fig. \ref{Fig:Kodak}\subref{Fig:BPP_Kodak} presents the R-D performance comparison under symmetric wireless conditions ($\eta_1=\eta_2=10$dB). We can observe that DARA-DJSCC still achieves superior R-D performance across the entire BPP range on the CLIC dataset. For instance, DARA-DJSCC achieves a PSNR of approximately 31.72dB at BPP=0.14, yielding a gain of about 0.55dB over h-DJSCC and 1.64dB over the traditional digital baseline. This improvement confirms the applicability and efficiency of DARA for high-resolution image transmission. Fig. \ref{Fig:Kodak}\subref{Fig:SNR_Kodak} further investigates hybrid transmission performance under varying wireless SNRs (with $\eta_1=10$dB and wired segment BPP fixed at 0.18). DARA-DJSCC exhibits a sustained advantage compared to other schemes, including h-DJSCC, modular transmission schemes and traditional QAM+LDPC combinations, showcasing its robust scalability for high-resolution data. The traditional digital baselines once again exhibit a sharp ``cliff effect'', in contrast to the graceful degradation demonstrated by DARA-DJSCC, showing its robustness for hybrid transmission of high-resolution contents.

\subsection{Visualization}
\begin{figure*}[!t]
\centering
\includegraphics[width=0.85\textwidth]{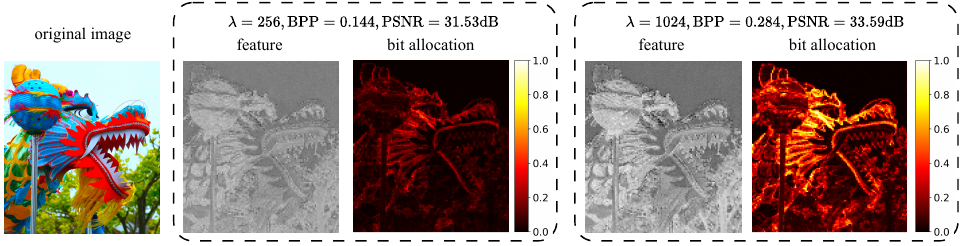}
\caption{Visualization of adapted features $\y_\lambda$ and bit
allocation heatmaps under different $\lambda$ values.}
\label{Fig:visualization}
\end{figure*}

Fig. \ref{Fig:visualization_ablation} presents the estimated distributions
of aligned features under different supervision strategies
at a wireless SNR of 10dB. The feature distribution obtained with contrastive supervision is closer to the teacher reference (denoted as ``Label'' in the figure) than those obtained with unsupervised training or MSE supervision, consistent with the reconstruction gains
reported in Table~\ref{tab:ablation}.

To intuitively demonstrate the effectiveness of the $\lambda$-conditioned variable-rate coding mechanism, we present a visualization of the adapted features and corresponding bit allocation heatmaps for a representative image from the CLIC dataset under different $\lambda$ values ($\lambda=256, 1024$ for lower and higher rate preference, respectively) in Fig. \ref{Fig:visualization}.

The visualized features illustrate how $\lambda$ influences the learned representations and corresponding bit allocation. For a given feature representation $\y$ derived from the DJSCC symbols, our proposed feature adaptor transforms it to $\yl$ before encoding, which is specifically tailored to the current coding preference dictated by the Lagrangian multiplier $\lambda$. When the rate constraint is strict, corresponding to a lower $\lambda$ value, the adapted features tend to retain only the fundamental and most essential information for basic image reconstruction. Conversely, when more bits are permitted for transmission, as indicated by a higher $\lambda$ value, the adapted features $\yl$ preserve richer textural details and exhibit a clearer distinction from the background, leading to superior reconstruction quality. Correspondingly, since texture and detailed regions typically possess higher entropy, they are allocated a greater number of bits for encoding.

By continuously varying $\lambda$, a smooth transition from coarse to fine feature representations can be achieved. This continuous adaptability is highly beneficial for meeting the dynamic and fluctuating transmission rate constraints encountered in practical communication scenarios, which demonstrates the effectiveness of our coding mechanisms. 

\subsection{Evaluation of Rate Control}
\begin{table*}[!htbp]
    \begin{center}
    \caption{Average rate control results on CIFAR-10 dataset}
    \label{tab:rate_control}
    \begin{tabular}{cccccccccc}
        \toprule
        \multirow{2}{*}{} & \multicolumn{3}{c}{$R_c=0.6\text{bpp}$} & \multicolumn{3}{c}{$R_c=0.9\text{bpp}$} & \multicolumn{3}{c}{$R_c=1.2\text{bpp}$}\\
        \cmidrule(r){2-4}
        \cmidrule(r){5-7}
        \cmidrule(r){8-10}
          & $2\text{dB}$ & $5\text{dB}$ & $10\text{dB}$ & $2\text{dB}$ & $5\text{dB}$ & $10\text{dB}$ & $2\text{dB}$ & $5\text{dB}$ & $10\text{dB}$\\
        \midrule
        $\lambda$ & 601.86 & 511.96 & 452.06 & 1825.74 & 1524.14 & 1321.52 & 3871.12 & 3401.39 & 3068.27 \\
        \midrule
        $R_a(\text{bpp})$ & 0.610 & 0.606 & 0.604 & 0.904 & 0.903 & 0.902 & 1.170 & 1.182 & 1.190 \\
        \midrule
        Relative error ($\frac{\left | R_c-R_a \right | }{R_c}$) & 1.67\% & 1.00\% & 0.67\% & 0.44\% & 0.33\% & 0.22\% & 2.5\% & 1.5\% & 0.83\% \\ 
        \bottomrule
    \end{tabular}
    \end{center}
\end{table*}
In this subsection, we evaluate the accuracy of the proposed Lagrangian multiplier-based rate control.

As a novel module, the Lagrangian multiplier predictor aims to model the mapping between the rate $R$ and the Lagrangian multiplier $\lambda$. Based on this module, an appropriate $\lambda$ for the wired rate constraint $R_c$ can be predicted by LMP. Table \ref{tab:rate_control} provides concrete numerical evidence for the rate control precision across various SNR and target rate combinations. To control the wireless channel conditions, the two wireless channels have the same SNR. It demonstrates that the actual rates ($R_a$) based on predicted $\lambda$ consistently align very closely with the rate constraints ($R_c$). For instance, at SNR=10dB and a rate constraint of 0.6 bpp, DARA achieves an actual rate of 0.604 bpp, with a relative error of only 0.67\%. Similarly, for a rate constraint of 0.9 bpp at SNR=5dB, the actual rate is 0.903 bpp, with a relative error of just 0.33\%. These results quantify the average deviation from target rates under various wireless conditions. Furthermore, Table \ref{tab:rate_control} confirms that the predicted $\lambda$ values, which control coding preference, consistently increase with the target rate, enabling the model to achieve higher rates for improving reconstruction quality. However, in gain-based schemes, each predefined discrete level corresponds to a fixed $\lambda$, but because image complexity varies, applying the same gain results in unpredictable rates. Since wired constraints are unknown during training, these discrete methods often require a trial-and-error search among all pretrained $\lambda$ values, leading to significant rate mismatches.

For the same target rate, Table~\ref{tab:rate_control} shows that the predicted
$\lambda$ decreases as the wireless SNR increases.
This trend may reflect SNR-dependent changes in the aligned
features $\y$ and their compressibility. It also indicates
that the $R$--$\lambda$ relationship depends on the received
representation, supporting the use of $\y$ as an input
to the predictor.

\section{Conclusion} \label{Section:conclusion}
This paper introduces DARA to adapt DJSCC symbols to wired links, improving E2E performance of DJSCC systems operating over hybrid wireless-wired networks. To resolve the channel quality-induced changes of symbol distributions, DARA employs contrastive supervision to align the
distribution of features derived from received DJSCC symbols with the latent feature distribution of a pre-trained NIC encoder, improving entropy coding
efficiency for wired forwarding. Furthermore, we leverage the Lagrangian multiplier method to achieve controllable and continuous variable-rate coding within a single model, thereby maximizing wired resource utilization under dynamic bandwidth constraints. Extensive experiments on CIFAR-10 and CLIC datasets validate the effectiveness of DARA, which demonstrates superior R-D performance and robust adaptation across diverse wireless and wired conditions compared to existing baselines. Extension of the proposed framework to more complex data modalities, e.g., videos, is an important direction for future work.

\end{document}